\documentclass[10pt,conference,letterpaper]{IEEEtran}
\usepackage{amsmath}
\usepackage[linesnumbered,ruled,vlined]{algorithm2e}
\usepackage{graphicx}
\usepackage{pifont}
\usepackage{tikz,pgfplots}
\usetikzlibrary{arrows,backgrounds,snakes}
\usetikzlibrary{patterns}
\usepackage{cite}
\usepackage{subcaption}
\usepackage{multicol}
\usetikzlibrary{calc}
\usepackage{comment}
\usepackage{mathtools}
\usepackage{array}
\usepackage{eqparbox}
\usepackage{gnuplottex}
\usepackage{color, colortbl}
\usepackage{booktabs}
\usepackage{amsthm}
\usepackage{paralist}
\usepackage{url}

\input{savespace.tex}

\newcommand{\fullversion}[1]{#1}
\newcommand{\confversion}[1]{}

\newcommand{\revise}[1]{#1}
\newcommand{\mycomment}[1]{}
\newcommand{\mysubsection}[1]{\vspace{1ex}\noindent\textbf{#1.}\xspace}

\newcommand{\etal}{\textit{et al.}\xspace}

\newcommand{\topk}{top-$k$\xspace}

\newcommand{\knn}{$k$NN\xspace}

\newcommand{\size}[1]{|#1|}

\newcommand{\union}{\cup}

\newcommand{\conj}{\wedge}

\newcommand{\twoldots}{\,\ldotp\ldotp\xspace}

\newcommand{\set}[1]{\{\, #1 \,\}}
\newcommand{\pair}[1]{\langle #1 \rangle}

\newcommand{\pexeso}{\textsf{PEXESO}\xspace}
\newcommand{\pexfor}{\textsf{PEXESO}\xspace}
\newcommand{\ept}{\textsf{EPT}\xspace}
\newcommand{\eptfor}{\textsf{EPT}\xspace}
\newcommand{\hg}{\textsf{PEXESO-H}\xspace}
\newcommand{\hgfor}{\textsf{PEXESO-H}\xspace}
\newcommand{\pq}{\textsf{PQ}\xspace}
\newcommand{\ctree}{\textsf{CTREE}\xspace}
\newcommand{\ctreefor}{\textsf{CTREE}\xspace}

\newtheorem{definition}{Definition}

\newtheorem{lemma}{Lemma}

\newcommand{\verify}{\textsf{Verify}\xspace}
\newcommand{\block}{\textsf{Block}\xspace}

\definecolor{Gray}{gray}{0.9}
\newcolumntype{g}{>{\columncolor{Gray}}c}

\SetFuncSty{text}
\SetArgSty{text}
\SetKw{OR}{or}
\SetKw{AND}{and}
\SetKw{NOT}{not}
\SetKw{TRUE}{true}
\SetKw{FALSE}{false}
\SetKw{NULL}{nil}
\SetKw{Break}{break}
\SetKw{Continue}{continue}
\SetKwInOut{Input}{Input}
\SetKwInOut{Output}{Output}
\newcommand{\State}[1]{#1\;}

\begin{document}


\title{Efficient Joinable Table Discovery in Data Lakes: A High-Dimensional Similarity-Based Approach}


%
%
%
%

\newcommand{\authorcop}{$^{\ast}$}
\newcommand{\authorsep}{\hspace{5mm}}
\newcommand{\authorone}{$^{\dagger}$}
\newcommand{\authortwo}{$^{\ddagger}$}
\newcommand{\authorthree}{$^{\S}$}
\newcommand{\authorfour}{$^{\natural}$}
\newcommand{\authorfive}{$^{\|}$}

\author{
    Yuyang Dong\authorone{} \authorsep{}
    Kunihiro Takeoka\authorone{} \authorsep{}
    Chuan Xiao\authortwo{} \authorsep{}
    Masafumi Oyamada\authorone{} \authorsep{}
   
    \vspace{1mm}\\
    {\fontsize{10}{10}\selectfont\itshape{\authorone{}NEC Corporation, Japan}}\hspace{1em}
    {\fontsize{10}{10}\selectfont\itshape{\authorone{}Osaka University and Nagoya University, Japan}}
    \\%
    {\fontsize{9}{9}\selectfont{\{dongyuyang, k\_takeoka, oyamada\}@nec.com}}\hspace{1em}
    {\fontsize{9}{9}\selectfont{chuanx@ist.osaka-u.ac.jp}}
    \\
}



\maketitle

Finding joinable tables in data lakes is key procedure in many applications such as 
data integration, data augmentation, data analysis, and data market. Traditional 
approaches that find equi-joinable tables are unable to deal with misspellings and 
different formats, nor do they capture any semantic joins. In this paper, we 
propose \pexeso, a framework for joinable table discovery in data lakes. \revise{We 
target the case when textual values are embedded as high-dimensional vectors and 
columns are joined upon similarity predicates on high-dimensional vectors}, hence to 
address the limitations of equi-join approaches and identify more meaningful results. 
To efficiently find joinable tables with similarity, we propose a block-and-verify 
method that utilizes pivot-based filtering. A partitioning technique is developed to 
cope with the case when the data lake is large and cannot fit in main memory. An 
experimental evaluation on real datasets shows that our solution identifies 
substantially more tables than equi-joins and outperforms other similarity-based 
options, and the join results are useful in data enrichment for machine learning tasks. 
The experiments also demonstrate the efficiency of the proposed method.

\section{Introduction} \label{sec:intro}
Join is a fundamental and essential operation that connects two or more tables. 
It is also a significant technique applied to relational database management 
systems and business intelligence tools for data analysis. The benefits of 
joining tables are not only in the database fields (e.g., data integration) 
but also in the machine learning (ML) fields such as feature augmentation and 
data enrichment~\cite{join-for-ml, arda}.

With the trends of open data movements by governments and the dissemination of 
data lake solutions in industries, we are provided with more opportunities to obtain 
a huge number of tables from data lakes (e.g., WDC Web Table Corpus \cite{URL:WDC}) 
and make use of them to enrich our local data. As such, a local table can be regarded 
as a query, and our request is to look for joinable tables in the data lake. Many 
researchers studied the problem of data discovery in data lakes. Unfortunately, 
existing works on finding joinable tables~\cite{josie, lshensemble} only focused on 
evaluating the joinability between columns by taking the overlap number of equi-joined 
records.
\revise{One example is joining the ``Race'' column in Table~\ref{tab:population} with the 
``Col\_1'' column in Table~\ref{tab:income}. ``White'' and ``Black'' yield equi-join 
results as strings exactly match in the two columns. However, the tables in data lakes 
usually do not have an explicitly specified schema and heterogeneous tables may differ in 
representations or terminologies, e.g., ``American Indian/Alaska Native'' v.s. ``Mainland 
Indigenous'', and ``Hawaiian/Guamanian/Samoan'' v.s. ``Pacific Islander''.} 
In these cases, equi-joins fail to capture the semantics. They either produce few join 
results if we use an inner-join, or cause sparsity if we use a left-join. This may not 
improve the effectiveness for ML tasks and sometimes even degrades the quality due to 
overfitting. On the other hand, despite a few recent studies on non-equi-joins, 
e.g., by string transformation~\cite{auto-join} or statistical correlation~\cite{sema-join}, 
they only deal with the case of joining two given tables; it is unknown how to find 
joinable tables in a data lake, and it is prohibitive to try joining every table in the 
data lake with the query table. Other recent advances, such as \cite{data-civilizer} and 
\cite{seeping-semantics}, can help users search for desired attributes in a data lake 
semantically, yet they do not consider if the identified columns are really joinable.

A deeper view on the semantic level, such as utilizing word embeddings, enables us to 
identify text with the same or similar meanings, hence to tackle the data heterogenity. 
We can solve the aforementioned drawbacks of equi-joins and cope with the joinable 
table search problem by representing each record of a column (in contrast to 
\cite{seeping-semantics} which uses word embeddings on column names) as a 
high-dimensional vector, and a column is thus represented as a multiset of high-dimensional 
vectors. Then, we can leverage the \emph{similarity} between vectors to evaluate the 
joinability between columns. 

\begin{table}[t]
  \caption{\revise{An example of semantically joinable tables.}}
  \begin{subtable}[t]{0.48\textwidth}
    \centering
    \caption{Population.}
    \begin{tabular}{ |l|r|r| } 
      \hline
      Race            & Population  & Median Age \\
      \hline
      White           & 234,370,202 & 42.0 \\ 
      Black           & 40,610,815  & 32.7 \\ 
      American Indian/Alaska Native & 2,632,102   & 31.7 \\ 
      Hawaiian/Guamanian/Samoan & 570,116 & 29.7 \\
      \hline
    \end{tabular}
    \label{tab:population}
  \end{subtable}
  
  \begin{subtable}[t]{0.48\textwidth}
    \centering
    \vspace{0.3cm}
    \caption{Median household income (US\$).}    
    \begin{tabular}{ |l|r| } 
      \hline
      Col\_1 & Col\_2 \\
      \hline
      White & 65,902 \\ 
      Black & 41,511 \\ 
      Mainland Indigenous & 44,772 \\ 
      Pacific Islander & 61,911 \\
      \hline
    \end{tabular}
    \label{tab:income}
  \end{subtable}
  \label{tab:simjoin}
\end{table}

In this paper, we study the problem of joinable table discovery in data lakes and explore in the 
direction of embedding records as high-dimensional vectors and joining upon similarity predicates. 
To the best of our knowledge, this is the first work targeting high-dimensional similarity on 
record embeddings for joinable table discovery. \revise{Some recent studies deal with the problem 
of joining tables for feature augmentation~\cite{join-for-ml, arda}, where a few candidate tables 
are assumed to be ready for join. They focused on efficient feature selection over these candidate 
tables; however, the assumption of the availability of candidate tables does not always 
hold, and our proposed solution can be used to feed them with such candidates.}

The problem of finding joinable tables with high-dimensional similarity has two challenges. 
First, the \emph{similarity computation} for high-dimensional data is expensive. For 
example, GloVe~\cite{URL:GloVe} transforms a word to a 50- to 300-dimensional vector. It is 
prohibitive to exhaustively compute the similarities between all pairs of records. 
Second, the \emph{number of tables} in a data lake is large. It is time-consuming to check 
whether the tables are joinable or not one by one. Existing research on high-dimensional 
similarity focused on searching an object or joining two datasets efficiently (see 
\cite{high-dimensional-tutorial} for a survey), 
but none of them were designed for searching for joinable table with similarity predicates. 

Seeing the above challenges, we propose a framework called 
\pexeso \footnote{PEXESO is a card game and the objective is to find matched pairs. https://en.wikipedia.org/wiki/Concentration\_(card\_game)}
to efficiently find joinable tables with high-dimensional similarity. \pexeso 
mainly deals with textual columns and support any similarity function in a metric 
space. The joinability of a table is measured by the number of matching records in 
the query column, which are defined using a distance function and a threshold. 
\pexeso adopts a block-and-verify strategy to reduce the similarity 
computation between records. \revise{We employ pivot-based filtering to select a 
set of pivot vectors and compute the distances to these pivots to prune vectors by 
the triangle inequality. Then hierarchical grids, which divide the pivot space into 
cells, are utilized to block vectors and find candidates. Finally, we verify the 
candidates to count the number of matching records with the help of an inverted index.} 
Our search algorithm finds exact answers to the joinable table search problem with 
similarity predicates. We analyze its complexity and cost. 
For the case of a large-scale data lake that cannot be loaded in main memory, we resort 
to data partitioning and load each part with a single \pexeso. We develop a clustering 
method that partitions the dataset by column distributions. 

We conduct experiments on real datasets and evaluate on ML tasks to show the 
effectiveness of our similarity-based approach of joinable table discovery as well as 
its usefulness in enriching data for ML. \pexeso achieves 0.21 -- 0.28 higher recall 
than the equi-join approach and outperforms the approaches using other similarity 
options such as Jaccard and fuzzy-join~\cite{fuzzytods} in both precision and recall. 
By using \pexeso for data enrichment, the performance of the ML tasks is improved by 
1.9\% higher micro-F1 score and 10\% lower mean squared error. 
As for efficiency, \pexeso outperforms exact baselines by up to 76 times speedup. 
Its processing speed is competitive with the approximate solution of product 
quantization~\cite{pq} (which has very low precision and recall in finding joinable 
tables) and even better in some cases. 

Our contributions are summarized as follows. 
\begin{inparaenum} [(1)]
  \item 
  We propose \pexeso, a framework for joinable search discovery in data lakes. Our 
  solution \revise{targets textual values embedded as high-dimensional vectors in a 
  metric space and columns are joined upon similarity predicates}. 
  \item 
  To efficiently find joinable tables upon similarity predicates, we design a block-and-verify solution based on 
  hierarchical grids and an inverted index. Our algorithm employs pivot-based 
  filtering to reduce similarity computation. 
  \item
  We propose data partitioning for the out-of-core case. A clustering method is 
  developed to partition the dataset. 
  \item
  We conduct experiments on real datasets to demonstrate the effectiveness and the efficiency of \pexeso in finding 
  joinable tables and its usefulness in building ML models. 
\end{inparaenum}

\revise{A caveat is that we do not target a specific representation learning
approach but efficiency optimization for the case when records are embedded in a metric space. 
Our design decouples effectiveness and efficiency optimizations. Any representation learning 
model can be used in our framework to transform the original data to vectors, as long as the 
output is in a metric space. The embedding approaches and the similarity we use 
in the experiments, according to the taxonomy in \cite{DBLP:conf/sigmod/MudgalLRDPKDAR18}, 
belong to the category of character-level pre-trained embedding, heuristic-based summation, 
and fixed distance comparison. As such, our solution renders the entire workflow unsupervised 
without any labelling work. We discover that this has already achieved better results than 
equi-join and non-semantic similarity approaches. We believe that more sophisticated models 
outlined in the design space~\cite{DBLP:conf/sigmod/MudgalLRDPKDAR18} may perform better, but 
this is specific to the task and may require labelled examples for training.} 

\fullversion{
The rest of the paper is organized as follows.  
Section~\ref{sec:framework} overviews the \pexeso framework and defines the joinable 
table search problem. Section~\ref{sec:algorithm} presents our indexing and search 
algorithm. Section~\ref{sec:boost} introduces data partitioning for large-scale 
data lakes. Section~\ref{sec:threshold} discusses threshold specification in \pexeso. 
Experimental results are reported and analyzed in Section~\ref{sec:exp}. 
Section~\ref{sec:related} reviews related work. Section~\ref{sec:concl} concludes the 
paper. 
}

\section{Joinable Table Discovery Framework}
\label{sec:framework}

\subsection{System Overview}

\begin{figure}[t]
  \centering
  \includegraphics[width=\linewidth]{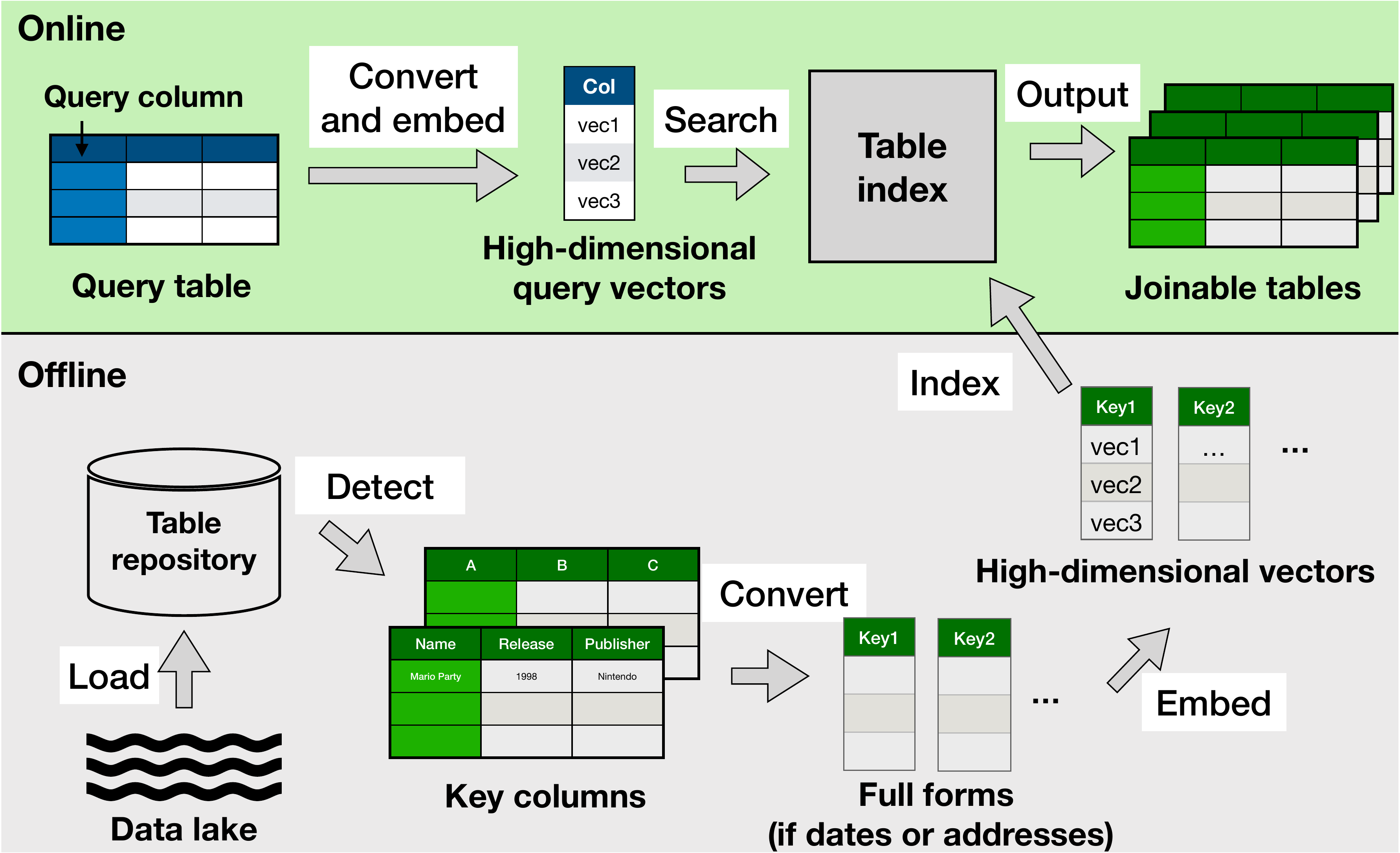}
  \caption{Joinable table discovery framework.}
  \label{fig:system}
\end{figure}

Fig.~\ref{fig:system} shows an overview of our \pexeso framework, which consists of two 
components:

\begin{itemize}
\item The offline component loads raw data (e.g., in CSV format) from the data lake to a 
table repository and extracts the columns that are expected to be join keys. 
For example, the WDC Web Table Corpus~\cite{URL:WDC} contains key column information. We may 
also use the SATO method~\cite{sato} to detect data types in tables and choose the columns whose types 
(e.g., names) can serve as a join key. For each string (including date) column, we transform 
the records (i.e., the string values rather than the column name) to high-dimensional vectors by 
a pre-trained model, e.g., fastText~\cite{URL:fasttext}, which carries semantic information and 
handles misspelling by making use of character-level information. In this sense, the pre-trained 
model can be regarded as a plug-in in our framework, and thus any representation learning method 
can be used here. \fullversion{For out-of-vocabulary words, the similarity can still be captured if 
we use subword embeddings. Another option is using the embedding of the most literally similar 
word or averaging the embeddings of the nearby context words.} 
To handle date and address columns, in which abbreviations often exist, we 
first convert abbreviations to their full forms (e.g., ``Mar'' to ``March'' and ``St'' to 
``Street'') and then apply the pre-trained model. 
\fullversion{When applying our method to domain-specific tables, we may leverage existing 
abbreviation dictionaries or knowledge bases for the domain, or learn a dictionary of 
abbreviation rules~\cite{DBLP:journals/pvldb/TaoDS17}.} 
The high-dimensional vectors are indexed for efficient lookup. 

\item The online component takes as input the user's query table, which contains a query column 
for join. There are several options to determine the query column: 
\begin{inparaenum} [(1)]
  \item the user specifies the query column, 
  \item we choose the string column with the most distinct values, and 
  \item we iterate through all the columns and regard each as a query column. 
\end{inparaenum}
Without loss of generality, we assume the first option, in line with \cite{josie}, and our 
techniques can be easily extended to support the other two options. We focus on the case of 
string columns as this is the most common data type in many data lakes (e.g., 65\% columns 
are strings in the WDC Web Table Corpus); for other data types, equi-join is used and this 
case has been addressed by Zhu \etal~\cite{josie}. The values in the query column are 
transformed to high-dimensional vectors using the same pre-trained model as in the 
offline component. Dates and addresses are also handled in the same way. Then we employ 
similarity predicates to define joinability and search for joinable tables in the repository. 
To present the results, we show the user a set of joinable tables along with the mapping 
between the records in the query column and the target column, since the user might not be 
familiar with our join predicates. 
\end{itemize}

\subsection{Similarity-Based Joinability}
Next we present a formal definition of the joinable table search problem. Table~\ref{tab:note} 
summarizes the notations frequently used in this paper. 

\begin{table}
  \small
  \centering
  \caption{Frequently used notations.}  
  \resizebox{\linewidth}{!}{%
  \begin{tabular}{|l|l|} \hline
    Symbol & Description\\ \hline \hline
    $Q, S$ & a query column, a target column in the repository \\ \hline    
    $R$ & a collection of target columns (i.e., the repository) \\ \hline
    $R_V$ & a collection of all the vectors in $R$'s columns \\ \hline
    $q, x$ & a query vector in $Q$, a target vector in the repository \\ \hline    
    $d(\cdot, \cdot)$ & a distance function  \\ \hline
    $\tau, T$ & a distance threshold, a column joinability threshold \\ \hline
    $M_\tau^d(a,b)$ & an indicator that indicates if $a$ matches $b$ \\ \hline
    $jn_\tau^d(Q, S)$ & the joinability of $S$ to $Q$ \\ \hline
    $p, P$ & a pivot vector, a set of pivot vectors \\ \hline
    $SQR(q', \tau)$ & a square query region  \\ \hline
    $RQR(q', p, \tau)$ & a rectangle query region  \\ \hline
    $HG_Q, HG_{R_V}$ & hierarchical grids for the mapped vectors of $Q$ and $R_V$ \\ \hline
    $m$ & the number of levels in a hierarchical grid\\ \hline
    $I$ & an inverted index \\ \hline
  \end{tabular}
  }
  \label{tab:note}
\end{table}

To match records at the semantic level, we consider vectors in a metric space and define the 
notion of vector matching under a similarity condition. 

\begin{definition} [Vector Matching]
  \label{def:vecmatch}
  Given two vectors $v_1$ and $v_2$ in a metric space, a distance function $d$, and a 
  threshold $\tau$, we say $v_1$ matches $v_2$, or vice versa, if and only if $d(v_1, v_2) \leq \tau$. 
\end{definition}

We use notation $M_\tau^d(v_1,v_2)$ to denote if $v_1$ matches $v_2$; i.e., 
$M_\tau^d(v_1,v_2) = 1$, iff. $d(v_1,v_2) \leq \tau$, or 0, otherwise. 

Given a query column $Q$ and a target column $S$, we use the number of matching vectors 
to define the joinability upon distance $d$ and threshold $\tau$, which counts the 
number of vectors in $Q$ having at least one matching vector in $S$, normalized by by 
the size of $Q$, i.e., 
\begin{align*}
  & jn_\tau^d(Q, S) = \frac{|Q_M|}{|Q|}, \\
  & Q_M = \set{q \mid q \in Q \conj \exists x \in S \text{\, s.t.\, } M_\tau^d(q, x) = 1}.
\end{align*}

Note the above joinability is not symmetric, i.e., we count matching vectors in $Q$ rather 
than $S$. We say the columns $Q$ and $S$ are \emph{joinable}, if and only if the joinability 
$jn_\tau^d(Q, S)$ is larger than or equal to a threshold $T$. We also say that the tables 
containing these two columns are \emph{joinable}. Next we define the joinable column (table) 
search problem. 

\begin{definition} [Joinable Column (Table) Search]
  \label{def:problem}
  Given a collection of columns $R$, a query column $Q$, a distance function $d$, 
  a distance threshold $\tau$, and a joinability threshold $T$, the joinable column (table) 
  search problem is to find all the columns in $R$ that are joinable to the query 
  column $Q$, i.e., $\set{S \mid S \in R \conj jn_\tau^d(Q, S) \geq T}$. 
\end{definition}

\revise{Duplicate values may exist in the query column. We regard them as independent 
records in the join predicate, because even if two records may share the same value in 
the query column, they may pertain to different entities.} 


\section{Indexing and Search Algorithm}
\label{sec:algorithm}
A naive method for the joinable table search problem is for each vector in $Q$, 
computing the distance to all the vectors in all the columns of $R$ 
and counting the number of matching vectors to determine if a column is joinable. 
The distance is computed $\size{Q} \cdot \sum_{S \in R} \size{S}$ 
times. Hence it is prohibitive when the number of vectors is large. To solve 
this problem efficiently, our key idea is to reduce the distance computation. 
We propose an algorithm that employs a block-and-verify strategy: the vectors of 
the query column and the target columns are blocked in hierarchical grids, and 
then candidates are produced by joining the cells in the hierarchical grids. We 
verify the candidates (i.e., to compute the exact distance between vectors) with 
the help of an inverted index while computing the joinabilities of the target 
columns. Our solution utilizes pivot-based filtering, which yields an exact answer 
of the problem. We do not choose approximate approaches to high-dimensional 
similarity query processing because they do not bear non-probability guarantee on 
the number of matching vectors and result in very low precision and recall (see 
Section~\ref{sec:effectiveness}). We begin with preliminaries on pivot-based filtering. 

\subsection{Preliminaries on Pivot-based Filtering}
The pivot-based filtering~\cite{pviotvldb17} uses pre-computed distances to prune 
vectors on the basis of the triangle inequality. The distance from each vector to 
a set of pivot vectors $P$ is pre-computed and stored. 
\confversion{
Then mismatched vectors can be pruned using the following lemma~\footnote{We provide the proofs 
in the extended version of the paper~\cite{fullversion}.}.}
\fullversion{
Then mismatched vectors can be pruned using the following lemma. 
}

\begin{lemma} [Pivot Filtering]
  \label{lem:pivot}
  Given two vectors $q$ and $x$, a set $P$ of pivot vectors, a distance function $d$, and 
  a threshold $\tau$, if $q$ matches $x$, then $d(q, p) - \tau \leq d(x, p) \leq d(q,p) + \tau$. 
\end{lemma}

\fullversion{
\begin{proof}
  We prove by contradiction.
  Assume that there exists an vector $x$ matches with $q$, i.e., $d(q, x) \leq \tau$,
  but for a pivot $p$, it holds $d(x,p) \notin [d(q,p) - \tau, d(q,p) + \tau]$, i.e., $|d(x,p) - d(q,p)| > \tau$.
  By the triangle inequality, $d(q,x) \geq |d(x,p) - d(q,p)| > \tau$, 
  and this contradicts the assumption. 
\end{proof}
}

Matching vectors can be identified using the following lemma. 

\begin{lemma} [Pivot Matching]
\label{lem:match}
  Given two vectors $q$ and $x$, a set $P$ of pivot vectors, a distance function $d$,
  and a threshold $\tau$, if there exists a pivot $p \in P$ such that $d(x, p) + d(q, p) \leq \tau$, 
  then $q$ matches $x$. 
\end{lemma}

\fullversion{
\begin{proof}
  For a pivot $p$, the triangle inequality holds: $d(q,p) + d(x,p) \geq d(q,x)$.
  If $d(x,p) \leq \tau - d(q,p)$, then $d(x,p) + d(q,p) \leq \tau$.
  Therefore, $d(q,x) < \tau$ and $x$ is matched with $q$.
\end{proof}
}

To utilize the above lemmata, pivot mapping was introduced~\cite{pviotvldb17}. 
Given a set of pivots $P = \{p_1, p_2, ..., p_n\}$, the pivot mapping for a vector $x$ 
involves computing the distance between $x$ and all the pivots in $P$, and assembling 
these values in a \emph{mapped} vector $x'$. Specifically, $x$ is mapped to the pivot space of 
$P$ as $x' = [d(p_1, x), d(p_2,x), ..., d(p_n, x)]$. 
The pivot size should be smaller than the dimensionality of the original metric space, so 
that the dimensionality can be reduced through range query processing in the pivot space 
to avoid the curse of dimensionality. Next we use an example to illustrate how to reduce 
distance computation by pivot mapping.

\begin{figure}[t]
  \centering
  \includegraphics[width=\linewidth]{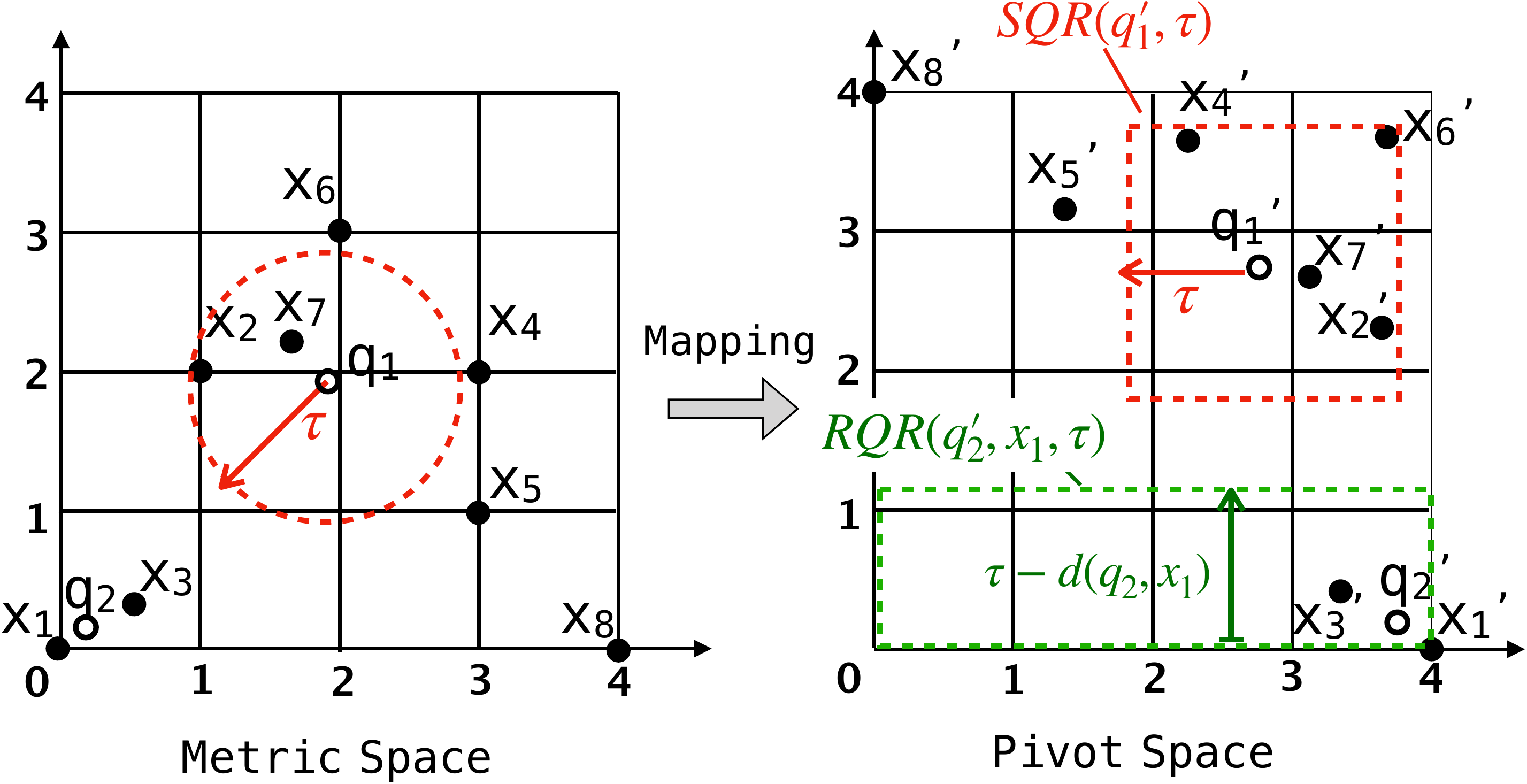}
  \caption{Example of pivot mapping with pivots $x_1$ and $x_8$,
  a square query region of pivot filtering for $q_1$ (red),
  and a rectangle query region of pivot matching for $q_2$ (green). 
  }
  \label{fig:pivotmap}
\end{figure}

Fig.~\ref{fig:pivotmap} shows an example in a 2-$d$ metric space. A query column $Q$ 
has two vectors: $Q = \set{q_1, q_2}$. There are four target columns in the table 
repository, each of them having two vectors: 
$S_1:\set{x_1, x_2}$, $S_2:\set{x_3, x_4}$, $S_3:\set{x_5, x_6}$, $S_4:\set{x_7, x_8}$.
These vectors are represented as points in a 2-$d$ metric space. Suppose $x_1$ and 
$x_8$ are selected as pivots and all the vectors are mapped to a 2-$d$ pivot space. 
In the pivot space, for each query vector $q$ and the distance threshold $\tau$, a 
square query region $SQR(q', \tau)$ is created, with $q$'s mapped vector $q'$ as the 
center and $2\tau$ as edge length. By Lemma~\ref{lem:pivot}, all the vectors outside 
the square query region $SQR(q', \tau)$ can be safely pruned from the result of the 
range search in the original metric space; i.e., none of them matches $q$. Therefore, 
only the vectors located in $SQR(q', \tau)$ need to be computed if they match $q$ via 
distance computation. In Fig.~\ref{fig:pivotmap}, only $x_2, x_4, x_6, x_7$, whose 
mapped vectors are located in the square query region $SQR(q_1', \tau)$ (in red), 
need distance computation against $q_1$. 

To use Lemma~\ref{lem:match} and find matching vectors, for each query $q$ and each 
pivot $p_i \in P$, a rectangle query region $RQR(q', p_i, \tau)$ is created. It  
starts from the original point $(0, 0)$; the edge length in the $i$-th dimension is 
$\tau - d(q, p)$, and the other edges have an infinite length. An exception is that 
the rectangle query region is not created for pivot $p_i$ when $\tau - d(q, p)$ is 
negative. By Lemma~\ref{lem:match}, all the vectors in $RQR(q', p, \tau)$ match query 
$q$. In Fig.~\ref{fig:pivotmap}, $q_1'$ has no rectangle query region for pivot 
$x_1$ or $x_8$ (due to negative edge length), and $q_2'$ has a rectangle query region 
$RQR(q_2', x_1, \tau)$ (in green) for pivot $x_1$, denoted as. Because $x_3'$ is 
located in this region, $x_3$ is guaranteed to match $q_2$ and thus there is no need 
to compute distance for them.

\begin{figure}[t]
  \centering
  \includegraphics[width=\linewidth]{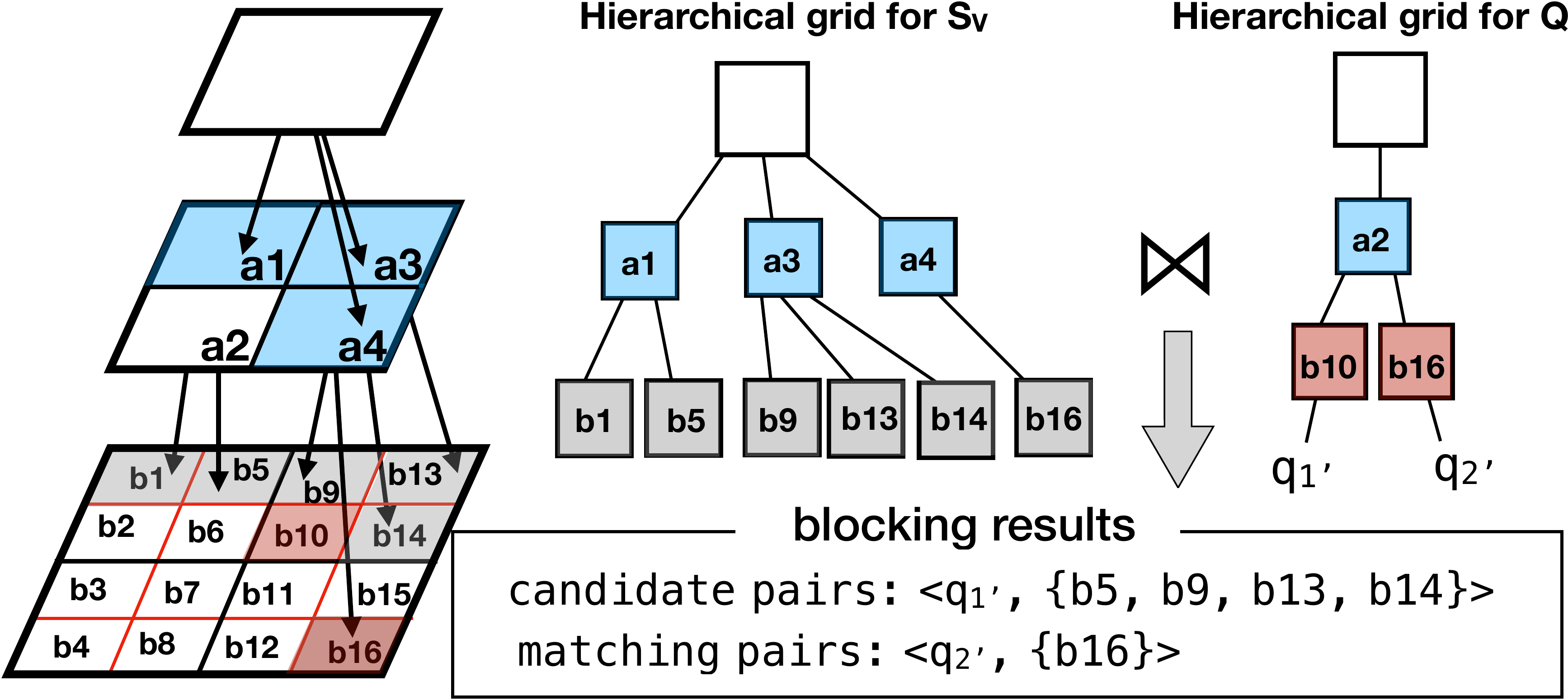}
  \caption{Hierarchical grids for target vectors and query vectors, and the 
  blocking results of matching pairs and candidate pairs.}
  \label{fig:hgrid}
\end{figure}
\label{sec:hgrid}

\subsection{Blocking with Hierarchical Grids}
Using the above techniques, we still need to check target vectors (i.e., the vectors 
in the columns of the table repository) against the query region of each query vector. 
To remedy this, we group vectors and employ pivot-based filtering to prune in a 
group-group manner, so the comparison between vectors and query regions can be 
significantly reduced. 

To achieve this, we propose to group similar mapped vectors. The pivot space is 
equally partitioned into small (hyper-) cells, and we manage the cells in a 
hierarchical grid with multiple levels of different partitioning granularity. We 
consider a hierarchical grid of $m$ levels (except the root) and divide the pivot 
space into $2^{|P|\cdot i}$ partitions, where $|P|$ is the dimensionality of the 
pivot space and $i \in [1 \twoldots m]$ is the level number. Fig.~\ref{fig:hgrid} 
shows an example of a 2-level hierarchical grid for the mapped vectors of 
$R_V$ in Fig.~\ref{fig:pivotmap}. For the 2-$d$ pivot space, we have two 
levels with $2^{2 \times 1} = 4$ cells and $2^{2 \times 2} = 16$ cells. Note that to 
save memory, the hierarchical grid only indexes the cells that have at least one 
vector. As such, in Fig.~\ref{fig:hgrid}, the leaf level has cells $b_1$, $b_5$, 
$b_9$, $b_{13}$, $b_{14}$ and $b_{16}$, and the intermediate level has cells $a_1$, 
$a_3$ and $a_4$.

Based on the above grouping strategy, Lemma~\ref{lem:pivot} yields the following 
filtering principle. 

\begin{lemma} [Vector-Cell Filtering]
  \label{lem:vcfilter}
  Given a cell $c$ and a mapped query vector $q'$ in the pivot space, 
  if $c \cap SQR(q', \tau) = \emptyset$, 
  then for any mapped vector $x' \in c$, its original vector $x$ does not match the query vector $q$. 
\end{lemma}
\fullversion{
\begin{proof}
$c \cap SQR(q', \tau) = \emptyset$ means none of the vectors in $c$ is located in $SQR(q', \tau)$. 
For each mapped vector $x' \in c$, its original vector $x$ satisfies that 
$d(x, p) \notin [d(q, p) - \tau, d(q, p) + \tau], \forall p \in P$. Hence by Lemma~\ref{lem:pivot}, 
$x$ does not match $q$. 
\end{proof}
}

We can also group the mapped query vectors into cells and compute a square query region 
for each cell $c_q$ as $SQR(c_q.center, \tau + \frac{c_q.length}{2})$, where $c_q.center$ 
is the center of the cell and $c_q.length$ is the edge length of the cell. To differentiate 
the two types of cells, the cells in the hierarchical grid for $R_V$ are 
\emph{target cells}, and those in the hierarchical grid for $Q$ are \emph{query cells}. 
Then, by Lemma~\ref{lem:pivot} we have 

\begin{lemma} [Cell-Cell Filtering]
\label{lem:ccfilter}
  Given a target cell $c$ and a query cell $c_q$ in the pivot space, 
  if $c \cap SQR(c_q.center, \tau + \frac{c_q.length}{2}) = \emptyset$, 
  then for any mapped vector $x' \in c$ and any query vector $q' \in c_q$, 
  their original vectors do not match. 
\end{lemma}

\fullversion{
\begin{proof}
For any vector outside $SQR(c_q.center, \tau + \frac{c_q.length}{2})$, 
its original vector $x$ satisfies that $d(x, c_q.center) > \tau + \frac{c_q.length}{2}$, 
and for each $q' \in c_q$, its original vector satisfies that $d(q, c_q.center) < \frac{c_q.length}{2}$.
Therefore, for any original vector $x$ and query vector $q$, $d(x, q) > \tau$. 
\end{proof}
}

Similar to the above strategy, we also extend Lemma~\ref{lem:match} to vector-cell
matching and cell-cell matching: 

\begin{lemma} [Vector-Cell Matching]
\label{lem:vcmatch}
  Given a target cell $c$ and a mapped query vector $q'$ in the pivot space, 
  if there exists a pivot $p \in P$ such that $c \cap RQR(q', p, \tau) = c$,
  then for any vector $x' \in c$, the original vector $x$ matches the query vector $q$.
\end{lemma}
\fullversion{
\begin{proof}
  For pivot $p \in P$,
  $c \cap RQR(q', p, \tau) = c$ means all the vectors in $c$ are inside the region $RQR(q', p, \tau)$,
  and for each vector $x' \in c$, its original vector $x$ satisfies that $d(x, p) \leq \tau - d(q, p)$.
  By Lemma~\ref{lem:match}, $x$ matches $q$. 
\end{proof}
}

For each pivot $p \in P$, we define the minimum rectangle query region as the intersection 
of all the $RQR(\cdot, \cdot, \cdot)$ of all the mapped query vectors in $c_q$, and denote 
the minimum rectangle query region as $\min(RQR(q', p, \tau)), q' \in c_q$. If any mapped 
query vector $q' \in c_q$ does not have a rectangle query region with a pivot $p$ due to 
negative edge length, then we define $\min(RQR(q', p, \tau))$ as an empty region. 

\begin{lemma} [Cell-Cell Matching]
\label{lem:ccmatch}
  Given a target cell $c$ and a query cell $c_q$ in the pivot space, if there exists a pivot 
  $p \in P$ such that $c \cap \min(RQR(q', p, \tau)) = c$, 
  then for any mapped vector $x' \in c$ and any query vector $q' \in c_q$, their original vectors match. 
\end{lemma}
\fullversion{
\begin{proof}
For a pivot $p \in P$, if $\min(RQR(q', p, \tau))$ exists,
then all the query vectors has rectangle query regions for $p$.
Therefore, if $c \cap \min(RQR(q', p, \tau)) = c$, $c$ must be covered by all of these rectangle query regions.
By Lemma~\ref{lem:match}, for any mapped vector $x' \in c$ and any query vector $q' \in c_q$, 
their original vectors satisfy $d(x, q) \leq \tau$. 
\end{proof}
}

We index the mapped vectors of $Q$ and $R_V$ in two hierarchical grids 
$HG_{Q}$ and $HG_{R_V}$. They slightly differ in structure: $HG_{Q}$ 
associates the mapped vectors of $Q$ in its leaf cells, but $HG_{R_V}$ 
does not (see Fig.~\ref{fig:hgrid}). 
The reason for such design is because the blocking phase aims to find pairs in the 
form of $\pair{mapped~query~vector, leaf~cells}$. There are two kinds of pairs 
found: matching and candidate pairs. Matching pairs are vector-cell pairs that 
satisfy Lemma \ref{lem:vcmatch}. Candidate pairs are pairs that cannot be filtered 
by Lemma~\ref{lem:vcfilter} or Lemma~\ref{lem:ccfilter}. In Fig.~\ref{fig:hgrid}, 
the blocking result is $\pair{q_2', \set{b16}}$ for matching pairs 
and $\pair{q_1', \set{b5, b9, b13,b14}}$ for candidate pairs. We use the form of 
$\pair{mapped~query~vector, leaf~cells}$ because the vectors in different 
columns of $R$ may share a common leaf cell in $HG_{R_V}$. 
Pairing leaf cells (instead of vectors or columns) with mapped query vectors 
exploits such share and yields efficient verification, as will be introduced later. 

\begin{algorithm}[t]
  \small
  \Input{parent cell $C_Q$, parent cell $C_R$, matching pair set $mPair$, candidate pair set $cPair$}
    \ForEach{child $c_Q \in C_Q$}{  
        \ForEach{child cell $c_R \in C_R$}{
            \If{$c_Q$ and $c_R$ are leaf cells}{
                \ForEach{vector $q' \in c_Q$}{
                    \If{$q$ and $c_R$ are matched by Lemma \ref{lem:vcmatch}}{
                        \State{$mPair \gets mPair \union \set{q', \set{c_R}}$ } 
                    }\Else{
                        \If{$q'$ and $c_R$ are not filtered by Lemma \ref{lem:vcfilter}}{
                            \State{$cPair \gets cPair \union \set{q', \set{c_R}}$ } 
                        }
                    }
                }
            }\Else{
                \If{$c_Q$ and $c_R$ are matched by Lemma~\ref{lem:ccmatch}}{
                        \State{$mPair \gets mPair \union \pair{q', \set{c}}$, for each vector $q' \in c_Q$ and leaf cell $c \in c_R$ } 
                }\Else{
                    \If{$c_Q$ and $c_R$ are not filtered by Lemma~\ref{lem:ccfilter}}{
                        \State{$\block(c_Q, c_R, mPair, cPair)$} 
                    }
                }
            }
        }
    }
  \caption{$\block(C_Q, C_R, mPair, cPair)$}
  \label{alg:block}
\end{algorithm}

To retrieve matching and candidate pairs efficiently, $HG_{Q}$ and $HG_{R_V}$ 
are constructed with the same number of levels. We propose an algorithm (Algorithm \ref{alg:block}) 
which follows a block nested loop join style but in a hierarchical way and scans 
$HG_{Q}$ and $HG_{R_V}$ only once. In particular, cells in $HG_{R_V}$ 
are pruned with the same level cells in $HG_{Q}$, and the 
sub-cells (i.e., children) are expanded at the same time on two hierarchical grids.
We use Lemmata \ref{lem:ccfilter} and \ref{lem:ccmatch} to filter and match non-leaf cells,
and use Lemmata \ref{lem:vcfilter} and \ref{lem:vcmatch} to filter and match leaf cells. 
Finally, the pairs of query vectors and corresponding leaf cells in $HG_{R_V}$ 
are retrieved as either candidate or matching pairs.

\subsection{Verifying with an Inverted Index}
\label{sec:invt}

After obtaining the matching pairs and candidate pairs, for each candidate pair 
$\pair{mapped~query~vector, leaf~cells}$, we compute the distances for the query vector 
and the target vectors in the leaf cells, and if they match, we increment the 
joinability count of the column having the target vector. We employ an inverted index 
in which the leaf cells of $HG_{R_{V}}$ are keys and each key corresponds to a 
postings list of columns associated with that key (i.e., having at least one vector in 
that cell). 

\begin{figure}[t]
  \centering
  \includegraphics[width=\linewidth]{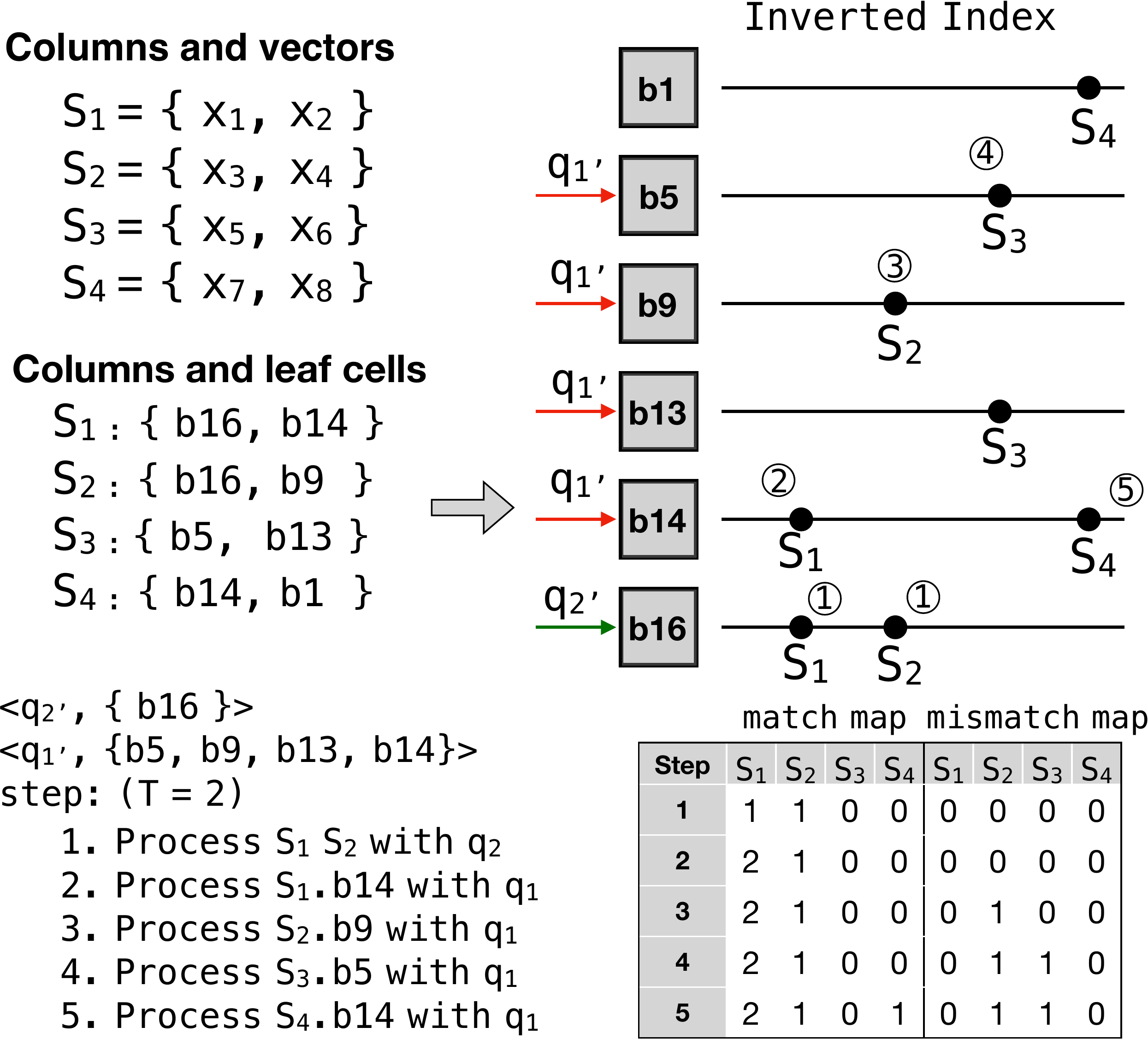}
  \caption{Inverted index for columns and leaf cells.}
  \label{fig:invt}
\end{figure}

Fig.~\ref{fig:invt} shows an example of the inverted index. We can look up the 
inverted index using the candidate and the matching pairs. We use two global maps to 
record two numbers: a \textit{match map} that records the number of matched vectors 
and an \textit{mismatch map} that records the number of mismatched vectors for each 
target column in the table repository. These recorded numbers are used to compute 
joinability for determining joinable columns.

For each matching pair, we increment the match map for the columns in the postings list. 
For each candidate pair, we look up the posting lists for the leaf cells in the 
candidate pair. During the lookup, we access the vectors indexed in the cell and use 
Lemmata \ref{lem:pivot} and \ref{lem:match} to filter and match these vectors. If any 
vector cannot be filtered or matched, we compute the exact distance to the query vector 
and update the match or mismatch map. Moreoever, we employ a DaaT (document-at-a-time 
\cite{daat}) paradigm for the inverted index lookup, where each column is regarded as a 
document. 
\revise{So columns in the inverted index are accessed by increasing order of ID. To 
implement this, we maintain a pointer for each postings list and 
pop the column with the smallest ID using a priority queue. For the sake of efficiency, 
we do not materialize a pointer for every cell but only those appear in the 
candidate set of the query vector.}
The benefit of the DaaT lookup is that it favors two early termination techniques: 
\begin{inparaenum} [(1)]
  \item whenever the joinability of a column exceeds the threshold $T$ during verification, 
  it is marked as joinable and we can skip processing any vector in this column, and 
  \item if a column has too many mismatched vectors and the remaining number of candidates 
  are not enough to make the matched vectors exceed $T$, we can early terminate the 
  verification of this column, as stated by the following filtering principle. 
\end{inparaenum}

\begin{lemma}
\label{lem:rfilter}
  Given a query column $Q$ and a target column $S$, let $U$ be any subset of $Q$ such 
  that none of the vectors in $U$ match any vector in $S$. If $|Q| - |U| < T$, then $S$ 
  is not a joinable column to $Q$.
\end{lemma} 
\fullversion{
\begin{proof}
    We prove by contradiction.
    Assume that $S$ is joinable to $Q$ and $|Q| - |U| < T$. Since there are at most 
    $|Q| - |U| < T$ matching vectors in $Q \backslash U$, there exists at least 
    one vector in $U$ such that the vector matches at least one vector in $S$. This 
    contradicts the definition of $U$. 
\end{proof}}

Fig.~\ref{fig:invt} shows the verification of matching pair $\pair{q_2', \set{b16}}$
and candidate pairs $\pair{q_1', \set{b5, b9, b13, b14}}$. Assume $T = 2$. 
In step 1, for $\pair{q_2', \set{b16}}$, 
we update $S_1$ and $S_2$ in the match map because they belong to the postings list of $b16$.
Then we process $\pair{q_1', \set{b5, b9, b13, b14}}$. 
In step 2, we check cell $b14$ of column $S_1$, denoted by $S_1.b14$. Since $S_1.b14$ has a 
vector $x_2$ and it matches $q_1$, $S_1$ in the match map is updated to 2. $S_1$ becomes 
a joinable column as the number of matching vectors reaches $T$. 
In step 3, \revise{by the DaaT lookup}, we reach $S_2.b9$, which has a vector $x_4$. Since 
it does not match $q_1$, $S_2$ in the mismatch map is updated to 1. In the same way, $S_3$ in 
the mismatch map is updated to 1 in step 4. After step 4, we do not need to check $S3.b13$ 
since the mismatch number for $S_3$ is 1, and $S_3$ can be filtered by Lemma~\ref{lem:rfilter}.
In step 5, we check $S4.b14$ and update the match map.  
The verification is finished and the result is $S_1$. 
Algorithm~\ref{alg:verify} gives the pseudocode of the verification. 

\begin{algorithm}[t]
  \small
  \Input{matching pair set $mPair$, candidate pair set $cPair$, inverted index $I$, thresholds $\tau$ and $T$}
  
  \ForEach{$\pair{q', \set{c}} \in mPair$}{
    \ForEach{leaf cell $c \in \set{c}$}{
      \State{Update the match map for the columns in $c$}
    }
  }
  
  \ForEach{$\pair{q', \set{c}} \in cPair$}{
    \ForEach{leaf cell $c \in \set{c}$}{
      \ForEach{column $S$ having at least one vector in $c$}{    
        \If {$S$ can be filtered by Lemma \ref{lem:rfilter}} {
           \Continue
        }
        \Else{
          \ForEach{vector $v' \in c$ that belongs to $S$}{
            \If{original vector $v$ of $v'$ is filtered by Lemma \ref{lem:pivot}}{
              \State{Update mismatch map for $S$}
            }
            \ElseIf{$v$ is matched by Lemma \ref{lem:match}}{
              \State{Update match map for $S$}
            }
            \Else{
              \State{Compute $d(q, v)$ and $M_{\tau}^d(q,v)$}
              \State{Update match map or mismatch map for $S$}
            }
            \If{$jn_{\tau}^d(Q, S) \geq T$}{
              \State{Mark $S$ as a joinable}
              \Continue
            }
          }
        }
      }
    }
  }
  \Return{the columns marked as joinable}
  \caption{$\verify(mPair, cPair, I, \tau, T)$}
  \label{alg:verify}
\end{algorithm}

\mysubsection{Quick browsing for inverted index}
Because we construct $HG_Q$ and $HG_{R_V}$ with the same level number, the leaf cells 
between them are also in the same granularity. If a query leaf cell and a target leaf cell 
refer to the same space region, then we can make sure that they cannot be filtered by 
Lemma~\ref{lem:vcfilter} or \ref{lem:ccfilter}. In this case, the query vectors and the target 
cell form candidate pairs. 
Therefore, we can get the leaf cells in $HG_Q$ and probe them in the inverted index directly. 
We call the above process \textit{quick browsing}. It processes some candidates in advance 
before running Algorithm \ref{alg:block}. 
Moreover, if quick browsing is issued, we can adjust Algorithm \ref{alg:block} 
easily for skipping the candidates in the same cell and avoiding redundant computations.

\begin{algorithm}[t]
  \small
  \Input{query column $Q$, hierarchical grid $HG_{R_V}$, inverted index $I$, thresholds $\tau$ and $T$}
  \Output{Joinable column set $J$}
  
  \State{Construct $HG_{Q}$}
  \State{Look up $I$ by quick browsing with $HG_{Q}.leafCell$ and update match and mismatch maps}
  \State{matching pair set $mPair\gets \emptyset$,  candidate pair set $cPair \gets \emptyset$}
  \State{$\block(HG_{Q}.root, HG_{R_V}.root, mPair, cPair)$}
  \State{$J \gets \verify(mPair, cPair, I, \tau, T$)}
  \Return{$J$}
  \caption{$\pexeso(Q, HG_{R_V}, I, \tau, T)$}
  \label{alg:pexeso}
\end{algorithm}

\subsection{Pivot Selection}
\label{sec:pivotselection}
The selection of pivots can significantly affect the performance of pivot filtering.
Good pivots can map original vectors and make them scattered in the pivot space, 
so as to make the query region cover fewer mapped vectors and filter more ones. 
Previous studies on pivot selection \cite{pviotvldb17, spbtreeTKDE17, pcapivot} have 
drawn the conclusion that good pivots are outliers but outliers are not always good 
pivots. Hence many methods~\cite{pviotvldb17, spbtreeTKDE17, pcapivot} pick outlier 
vectors as candidates and then select good pivots from them. Here, we adopt the 
PCA-based method~\cite{pcapivot} to select high quality pivots in $O(|R_V|)$-time.

\subsection{Search Algorithm and Complexity \& Cost Analysis}
\label{sec:analysis}

We assemble the above techniques and present an algorithm 
(Algorithm \ref{alg:pexeso}) for solving the joinable table search problem. 

\mysubsection{Time complexity}
The mapping and construction of the hierarchical grid for $Q$ has a complexity of 
$O((|P|+m)\cdot|Q|)$. 
The quick browsing and the block-and-verify method are $O(log|Q| \cdot log|R_V|)$.
The total time complexity of search is $O((|P|+m)\cdot|Q| + log|Q| \cdot log|R_V|)$.

For the construction of \pexeso, we used a PCA-based pivot selection algorithm with a complexity of $O(|R_V|)$. 
Pivot mapping takes $O(|P| \cdot |R_V|)$-time. 
For building the index, it takes $O(m \cdot |R_V|)$-time for 
the hierarchical grid and $O(D)$ for the inverted index, 
where $D$ is the total number of cells in all the columns.
The total time complexity of construction is $O((|P|+ m) \cdot |R_V| + D)$. 

Appending a new column $s$ into \pexeso takes $O(|P|+ m) \cdot |s|)$-time 
to pivot map $s$ and insert it into the corresponding cells of the hierarchical grid,
and it takes $O(1)$-time to insert $s$ into the corresponding postings lists of the inverted index.
Deleting a column $s$ from \pexeso takes $O(1)$-time to delete $s$ from the hierarchical grid,
and it takes $O(log|R|)$-time to locate and delete $s$ from the inverted index.


\mysubsection{Space complexity}
There are two hierarchical grids and an inverted index in \pexeso.
The space complexity for $HG_Q$ is $O(|Q|)$. For $HG_{R_V}$ and the inverted index, 
it is $O(|R_V| + D)$-space. The total space complexity is $O(|Q| + |R_V| + D)$.


\mysubsection{Cost analysis}
\label{sec:costmodel}
To estimate the cost of joinable table search with \pexeso, we analyze the 
expected number of distance computations for $d(\cdot, \cdot)$. Since 
blocking only compares overlap and does not compute $d(\cdot, \cdot)$, we 
only need to consider the cost in verification. Our experiment 
(Section~\ref{sec:para}) also shows that the blocking time is negligible in 
the entire search process.



Let $C$ denote the multiset of query vectors in the candidate pairs. The 
occurrence of a vector $q$ in $C$ is counted as the times it appears in 
the set of candidate pairs identified by the blocking. In verification, 
the expected number of distance computation is 
\begin{align}
  \label{equ:cost}
  E = \sum_{q \in C} N(SQR(q',\tau)), 
\end{align}
where $N(SQR(q',\tau)$ is the number of vectors in the leaf cells covered 
by the region of $SQR(q',\tau)$. Instead of estimating its exact value, we 
give an upper bound of $N(SQR(q',\tau)$. 
Assume the probability distribution function (PDF) for each dimension of 
the mapped vectors $R_V$ is $PDF_{i}(R_V), i \in [1,|P|]$. 
To obtain the vectors covered by $SQR(q',\tau)$, we need to take the 
intersection of vectors that cannot be filtered by any dimension of the pivot 
space. So the maximum number of the above 
intersection, denoted as $N_{\max}(SQR(q',\tau)$, is the minimum number of 
vectors in the covered region along all the dimensions of the pivot space. 
Thus we have 
\begin{align}
  \label{equ:cost-each-vector}
  N_{\max}(SQR(q',\tau)) = \min_{i \in [1,|P|]} ( \int_{q'[i] - \tau - \frac{1}{2^{|P|\cdot m}}}^{q'[i] + \tau + \frac{1}{2^{|P|\cdot m}}} PDF_{i}(R_V)). 
\end{align}

\mysubsection{Optimal $m$ for index construction}
Tuning $m$ is a trade-off between candidate number and inverted index 
lookup. To find an optimal $m$, we consider a query workload $\mathcal{Q}$: one 
option is to sample a subset of $R$ as query workload, and pair them with 
varying $\tau$ and $T$ values uniformly generated in a reasonable range for 
practical use (e.g, 0 -- 10\% maximum distance for $\tau$ and 20\% -- 80\% average 
column length for $T$, see Section~\ref{sec:threshold}). Then 
each query in the workload $\mathcal{Q}$ yields an estimated cost by Equation~\ref{equ:cost}. 
We can find an optimal $m$ by minimizing the overall expected 
cost across $\mathcal{Q}$ with an optimization algorithm such as gradient descent. 
To compute Equation (\ref{equ:cost}), we only do blocking to obtain $C$ but do not 
verify the candidates as this is very time-consuming for the entire query workload; 
instead, we estimate the cost for each query vector by Equation~\ref{equ:cost-each-vector}. 
In addition, since the value of $m$ obtained by gradient descent is fractional, we 
round by ceiling to get an integer value. 

\section{Partitioning for Large-Scale Datasets} 
\label{sec:boost}

\fullversion{
\begin{figure*}[t]
  \centering
  \begin{subfigure}[b]{0.3\textwidth}
  \centering
    \includegraphics[width=\linewidth]{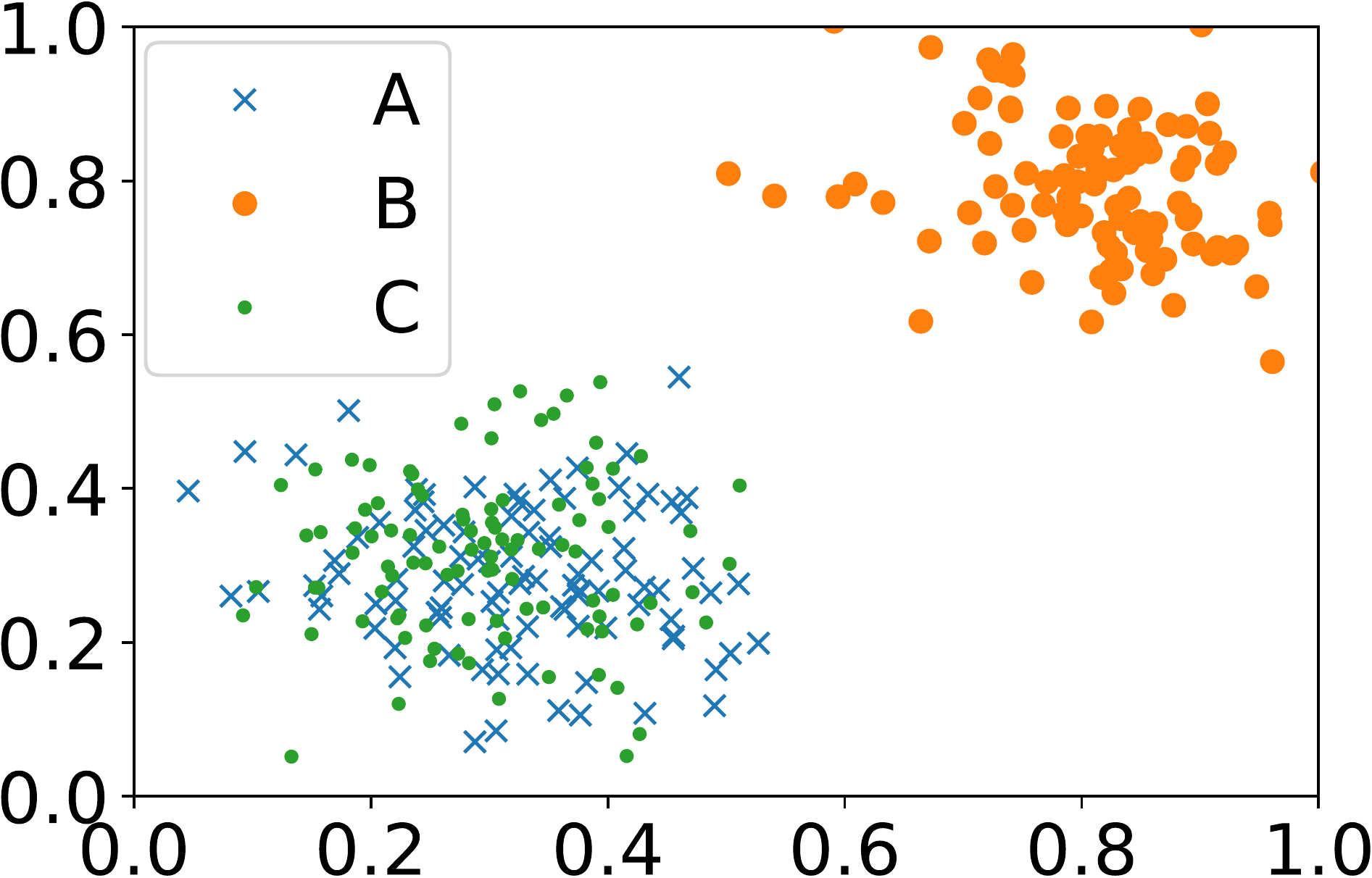}
    \caption{$A$, $B$, $C$ columns in original space.}
    \label{fig:abc:org}
  \end{subfigure}
   \begin{subfigure}[b]{0.3\textwidth}
   \centering
    \includegraphics[width=\linewidth]{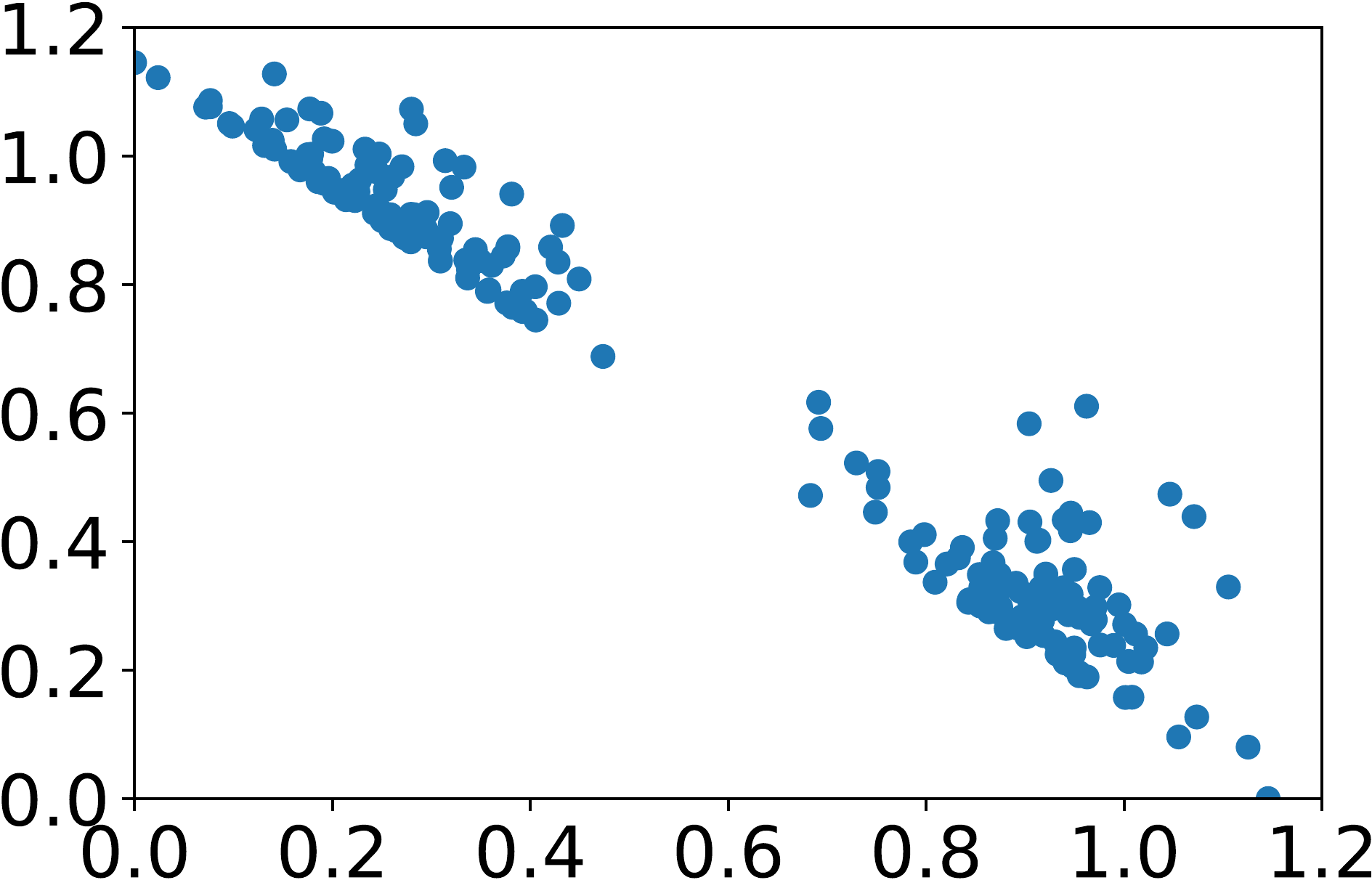}
    \caption{Pivot mapping $A$ and $B$ columns.}
    \label{fig:ab:map}
  \end{subfigure}
   \begin{subfigure}[b]{0.3\textwidth}
   \centering
    \includegraphics[width=\linewidth]{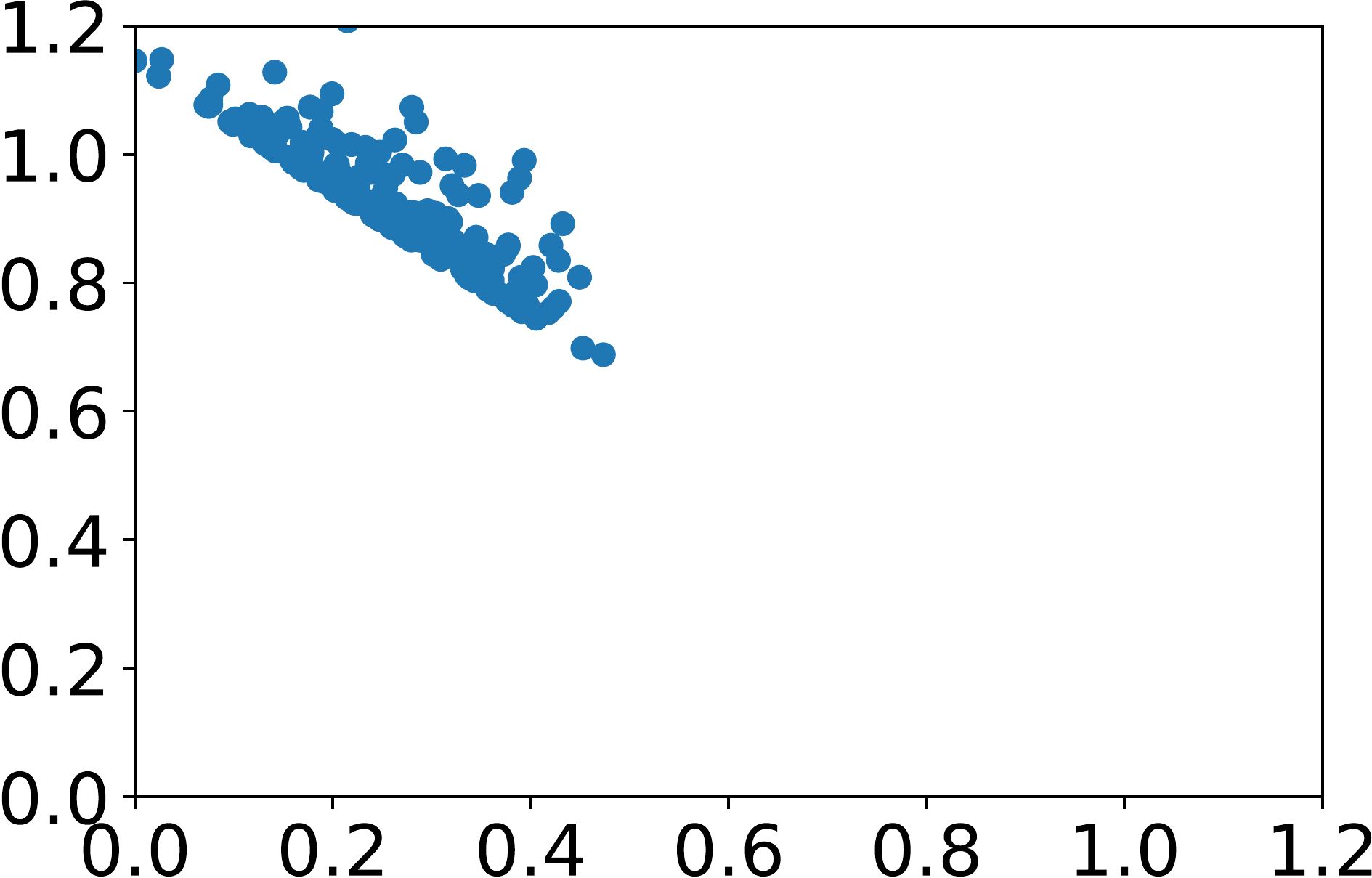}
    \caption{Pivot mapping $A$ and $C$ columns.}
    \label{fig:ac:map}
  \end{subfigure}
  \caption{Pivot mapping with different groupings.}
  \label{fig:abc}
\end{figure*}
}

A common scenario is that the number of columns extracted from the data lake is 
extremely large, and we cannot load all the data in a single \pexeso and hold them in 
main memory. Nonetheless, \pexeso is flexible in the sense that we can split the data 
into small partitions and each partition is indexed in a \pexeso framework. When 
processing a joinable table search, we load each partition into main memory at 
a time and search the results, and merge the results from from every \pexeso to obtain 
the final ones. 

An important problem is how to make a good partition that can maximizes the power of 
each \pexeso. To this end, we propose a data partitioning method based on a clustering 
with Jensen–Shannon divergence. 


Recall in Section~\ref{sec:pivotselection}, pivots are selected from outliers. 
One observation is that if we group columns with different data distributions, the power of 
selected pivots will decline. \confversion{We provide in the extended 
version~\cite{fullversion} an example.} 
\fullversion{For example, there are three columns $A$, $B$, and $C$ in Fig.~\ref{fig:abc:org}. 
$A$ has a similar distribution to $C$, and $B$ has a different distribution from them. 
If we group $A$ and $B$ together and select the pivots from them, 
the outliers in $A$ are far away from those in $B$, and the pre-computed distances 
will be not helpful to the pivot-based filtering of the vectors in $B$, and vice versa. 
On the other hand, if we group similar columns $A$ and $C$ together, the outliers in 
$A$ can also work for the vectors in $B$. 
Figs.~\ref{fig:ab:map} and \ref{fig:ac:map} are the pivot mapping result for 
$\set{A,B}$ and $\set{A,C}$, respectively. The white space is the filtered region. 
Obviously, grouping $\set{A, C}$ leads to better filter power than grouping $\set{A, B}$.}
Inspired by this observation, we choose to cluster the columns according to 
the similarity between distributions. KL divergence is a widely-used measure of such 
(dis)similarity. Since KL divergence is an asymmetric measure, we use the symmetric 
Jensen–Shannon divergence (JSD), a distance metric based on KL divergence. 

\begin{align*}
  JSD(A||B) = \frac{KLD(A||B) + KLD(B||A)}{2}, 
\end{align*}
where $KLD(A||B) = \sum_{x \in X} A(x) \cdot log(\frac{A(x)}{B(x)})$. 

We propose a clustering algorithm, which follows the $k$-means clustering 
paradigm. 
\begin{inparaenum} [(1)]
  \item Since JSD is a measure between probability distributions, we 
  summarize a column of vectors with a probability distribution histogram 
  composed of a number of bins, i.e., to obtain the statistics of the 
  probability of points in a space region.
  \item We randomly select $k$ columns as the center of $k$ clusters. 
  \item For each column (as a histogram), we compute the JSD distance to all 
  the $k$ centers, and assign this column to the cluster that yields the 
  minimum JSD. 
  \item For each cluster, we compute the mean of the histograms in this 
  cluster and update the center. 
  \item Steps (2) -- (4) are repeated until reaching a user-defined iteration 
  number $t$. 
\end{inparaenum}
The time complexity of the algorithm is $O(|R|\cdot k \cdot t)$. 



\section{Specifying Thresholds for Joinable Table Search}
\label{sec:threshold}
We discuss how to specify the two thresholds in our \pexeso framework. 
In general, we can convert the thresholds to ratios so users are able 
to specify them in an intuitive way, irrespective of data types, embedding 
approaches, or query column size. 

\begin{itemize}
    \item For distance threshold $\tau$, we first normalize all the 
    vectors to unit length. The maximum possible distance between any 
    two vectors is thus fixed (e.g., 2 for Euclidean distance). Then we 
    set the threshold as a percentage of the maximum distance: a larger 
    percentage indicates a looser matching condition and may increase 
    the number of retrieved joinable columns. 
    \item For joinability threshold $T$, we set it as a percentage of 
    the query column size. A larger percentage means a smaller number 
    of retrieved joinable columns.  
\end{itemize}

\section{Experiments} 
\label{sec:exp}

\subsection{Setting}
\confversion{We provide a summary here. 
Please refer to the extended version~\cite{fullversion} for detailed setup.} 

\mysubsection{Datasets}
We use the following datasets. Table~\ref{tab:datasets} summarizes the statistics, 
including the number of vectors, the number of string columns, the average number of 
vectors per column, the pre-trained model used to embed strings, and the dimensionality. 
\begin{inparaenum} [(1)]
    \item
        \textbf{OPEN} is a dataset of relational tables from Canadian Open Data Repository~\cite{lshensemble}. 
        We extract English tables that contain more than 10 rows. 
        \revise{We transform the values to 300-dimensional vectors with fastText~\cite{URL:fasttext}.}
    \item
        \textbf{WDC} is the WDC Web Table Corpus~\cite{URL:WDC}. 
        We use the English relational Web tables 2015. 
        \revise{String values are split into English words and GloVe~\cite{URL:GloVe} is used to 
        transform each word to a 50-dimensional vector. Then we compute the average of the word embeddings.} 
        We extract two subsets of this dataset, denoted by \textbf{SWDC} (small WDC, for in-memory) and 
        \textbf{LWDC} (large WDC, for out-of-core), respectively. 
\end{inparaenum}

For each dataset, we randomly sample 50 (for effectiveness) or 100 (for efficiency) 
tables from the dataset as query tables and removed it from the dataset to avoid 
duplicate result. We use Euclidean distance for the distance function $d(\cdot, \cdot)$. 
$\tau$ varies from 2\% to 8\% maximum distance (i.e., 2 for normalized vectors, see 
Section~\ref{sec:threshold}), and the default value is 6\%. 
$T$ varies from 20\% to 80\% query column size, and the default value is 60\%.

\fullversion{To construct the table repository, we load raw tabular data stored in CSV 
format and extract the columns that are possible to be join keys. The WDC dataset 
contains the key column information. For the datasets without such information, we 
use the SATO method~\cite{sato} to detect data types in tables and chose the 
columns whose types (e.g., names) can serve as a join key. Note that only string 
columns are involved in the evaluation because the other columns (e.g., IDs) in the 
datasets can be either dealt with equi-join, which has been addressed in \cite{josie}, 
or do not produce meaningful join results. We also remove tables that are vertical, lack 
key column information, or contain less than five rows.} 

\mysubsection{Competitors}
For effectiveness, we compare equi-join \cite{josie}, Jaccard-join (using Jaccard 
simialrity to match records), \revise{edit-join (using edit distance to match records)}, 
fuzzy-join~\cite{fuzzytods}, \revise{TF-IDF-join~\cite{DBLP:conf/sigmod/Cohen98}}, and our 
\pexeso. For efficiency, we consider the following competitors. 
\begin{inparaenum} [(1)]
  \item \textbf{\pexeso} is our proposed method. 
  
  \item \textbf{\hg} has the same hierarchical grid-based blocking as \pexeso but 
  replaces the inverted index-based verification with a naive method: for each candidate pair, 
  it computes the distance for the query vector and every vector in the cell. 
  
  \item \textbf{\ctree} is an exact method using cover tree~\cite{cover-tree-icml15}. 
  It builds a cover tree index for all the vectors and issues a range query with radius 
  $\tau$ for each vector in the query column. Then each result of the range query 
  is counted towards the joinability of the column it belongs to. 
  \revise{We use the implementation in \cite{URL:covertree}.}
  
  \item \textbf{\ept} is an exact method using a pivot table~\cite{ept}, which was suggested 
  in \cite{pviotvldb17} for its competitiveness in most cases. 
  It follows the same workflow as \ctree but replaces the cover tree with a pivot table. 
  \revise{We implement it by ourselves.} 
  
  \item \textbf{\pq} is an approximate method. 
  It follows the same workflow as \ctree but process the range query with product quantization~\cite{pq}. 
  \revise{We use the nanopq implementation~\cite{URL:nanopq}.}
  
\end{inparaenum}
\revise{Like \pexeso, we also equip the other methods with early termination in verification: 
when we increment the joinability counter of a column, if the count reaches $T$, then 
the column becomes joinable and we can skip it for further verification.} 

\begin{table}[t]
  \small
  \centering
  \caption{Dataset statistics.}  
  \resizebox{\linewidth}{!}{
  \begin{tabular}{ |l|c|c|c|c|l|l| } 
    \hline
    Dataset & \# Tab. & \# Vec. & \# Col. & Avg. Vec./Col. & Model & Dim. \\
    \hline
    \hline
    OPEN & 10.2K& 17.2M & 21.6K & 796 & fastText & 300\\
    SWDC & 516K& 8.6M & 516K & 16.7 & GloVe & 50\\  
    LWDC & 48.9M& 602M & 48.9M & 12.3 & GloVe & 50\\ 
    \hline
  \end{tabular}
  }
  \label{tab:datasets}
\end{table}

\begin{table}[t]
  \small
  \centering
  \caption{\revise{Precision \& recall of joinable table search.}}
  \scalebox{1}{
  \begin{tabular}{ |l|cc|cc| } 
    \hline
    & \multicolumn{2}{c}{OPEN} & \multicolumn{2}{c|}{SWDC} \\ \hline
    Methods & Precision & Recall & Precision & Recall \\
    \hline
    equi-join    & 1.000 & 0.611   & 1.000 & 0.589 \\
    Jaccard-join & 0.876 & 0.732   & 0.919 & 0.781 \\
    edit-join    & 0.814 & 0.756   & 0.833 & 0.842 \\
    fuzzy-join   & 0.834 & 0.795   & 0.865 & 0.833 \\
    TF-IDF-join   & 0.784 & 0.712   & 0.810 & 0.782 \\    
    \pexeso      & 0.911 & 0.821   & 0.948 & 0.868 \\
    our join with \pq-85 & 0.787 & 0.422   & 0.744 & 0.461 \\    
    \hline
    \end{tabular}
  }
  \label{tab:effect:joinable}
\end{table}

\mysubsection{Environments}
Experiments are run on a server with a 2.20GHz Intel Xeon CPU E7-8890 and 630 GB RAM.
\revise{All the competitors are implemented in Python 3.7.}


\begin{table*}[t]
  \centering
  \caption{Performance in ML tasks.}

  \begin{subtable}{0.33\textwidth}
    \centering
    \caption{Company classification.}    
    \scalebox{1}{
    \begin{tabular}{ |l|c|c| } 
      \hline
      Method & \# Match & Micro-F1 \\
      \hline
      no-join  & - & $0.825 \pm 0.057$ \\ 
      equi-join & 0.13\% & $0.806 \pm 0.069$ \\ 
      Jaccard-join & 0.54\% & $0.816 \pm 0.075$  \\
      fuzzy-join & 0.83\% & $0.836 \pm 0.083$  \\
      edit-join & 0.88\% & $0.831 \pm 0.063$ \\
      TF-IDF-join & 0.72\% & $0.826 \pm 0.083$ \\
      \pexeso  & 0.76\% & $\boldsymbol{0.855 \pm 0.045}$ \\ 
      \hline
    \end{tabular}
    }
    \label{tab:effect:company}
  \end{subtable}
  \begin{subtable}{0.33\textwidth}
    \centering
    \caption{\revise{Amazon toy product classification.}}
    \scalebox{1}{
    \begin{tabular}{ |l|c|c|c|c| } 
      \hline
      Method & \# Match & Micro-F1 \\
      \hline
      no-join       & -         & $0.589 \pm 0.077$ \\ 
      equi-join     & 0.09\%    & $0.586 \pm 0.051$ \\
      Jaccard-join  & 0.56\%    & $0.585 \pm 0.073$ \\
      fuzzy-join    & 0.83\%    & $0.592 \pm 0.055$ \\
      edit-join     & 0.89\%    & $0.596 \pm 0.067$ \\
      TF-IDF-join    & 0.82\%    & $0.594 \pm 0.049$ \\
      \pexeso       & 0.76\%    & $\boldsymbol{0.613 \pm 0.072}$ \\ 
      \hline
    \end{tabular}
    }
    \label{tab:effect:amazon}
  \end{subtable}
  \begin{subtable}{0.33\textwidth}
    \centering
    \caption{\revise{Video game sale regression.}}
    \scalebox{1}{
    \begin{tabular}{ |l|c|c| } 
      \hline
      Method & \# Match & MSE \\
      \hline
      no-join   & - & $2.10 \pm 0.75$ \\ 
      equi-join & 0.05\%       & $2.09 \pm 0.65$ \\
      Jaccard-join & 0.14\%    & $2.05 \pm 0.73$ \\
      fuzzy-join & 0.28\%      & $1.98 \pm 0.61$ \\
      edit-join & 0.53\%       & $2.01 \pm 0.83$ \\
      TF-IDF-join & 0.31\%      & $2.02 \pm 0.55$ \\
      \pexeso  & 0.64\% & $\boldsymbol{1.78 \pm 0.66}$ \\ 
      \hline
    \end{tabular}
    }
    \label{tab:effect:game}
  \end{subtable}
\end{table*}

\subsection{Effectiveness of Joinable Table Search}
\label{sec:effectiveness}
We first evaluate the effectiveness on OPEN and SWDC. 
We randomly sample 50 tables from each dataset as query table and specify 
a key column in each of them as query column. We request our colleagues of 
database researchers to label whether a retrieved table is joinable. 
Precision and recall are measured. 
Precision $=$ (\# retrieved joinable tables) $/$ (\# retrieved tables). 
Since it is too laborious to label every table in the dataset, we follow 
\cite{recall-search-engin} and build a retrieved pool using the union of the 
tables identified by the competitors: 
Recall $=$ (\# retrieved joinable tables) $/$ (\# joinable tables in the retrieved pool).

\revise{The thresholds of each competitor are tuned for highest F1 score.} 
Table~\ref{tab:effect:joinable} reports the average results. Equi-join has 100\% 
precision, but its recall is significantly lower than the other methods. Jaccard-join 
has higher precision than fuzzy-join but its recall is lower. \pexeso delivers the 
highest recall, and the advantage over equi-join and Jaccard-join is remarkable. 
\pexeso also outperforms Jaccard-join and fuzzy-join in precision, and achieves over 
90\% precision on both datasets. This showcases that \pexeso finds more joinable 
tables than other options and most of its identified tables are really joinable. 
Besides, in order to explain why choose an exact solution to the joinable table search 
problem, we replace our algorithm with an approximate method of product quantization 
to find matching vectors and tune its recall of range query to 85\%. 
Such modification (dubbed ``our join with \pq-85'') results in low precision and 
recall, meaning that using an approximate solution to find matching vectors is not a 
good option. 

\subsection{Performance Gain in ML Tasks}
\label{sec:performance-gain}
We evaluate three ML tasks to show the usefulness of joinable table discovery. 
\revise{The columns are embedded using fastText \cite{URL:fasttext}. 
For the query table in each task, we sample 1,000 records and search for joinable 
tables in the SWDC dataset that serves as a data lake. Then we left-join the query 
table to the identified joinable tables. Due to the noise in SWDC, a column is 
discarded if the size (excluding missing values) is smaller than 200. 
\fullversion{
There are two types of possible conflicts in the join results: first, one record in 
the query column may match multiple records in a target column; second, since we 
join query table with all the identified joinable tables, these target tables may 
share the same column. The first type of conflict is not observed in this experiment. 
To address the second type, we aggregate the values of the columns with similar 
column names by string concatenation and summing up numerical values.} 
Recursive feature elimination is applied on the join results to select meaningful 
features. With these features, a random forest model is trained for testing the 
prediction accuracy. In addition to the aforementioned competitors, we also consider 
using the query table without joins (referred to as ``no-join''). 
We measure micro-F1 score for classification and mean squared error (MSE) for 
regression. Parameters are tuned to avoid overfitting. The best average scores of a 
4-fold cross-validation are reported.}

\mysubsection{Company classification}
We use the company information table~\cite{URL:company} which contains 73,935 companies 
with 13 classes of categories (professional services, healthcare, etc.).
The task is to predict the category of each company. We use ``company\_name'' as query column.
Table~\ref{tab:effect:company} reports the average micro-F1 score and the number of records in 
the data lake identified as match to those in the query table. Equi-join only finds 0.13\% 
matching records in the data lake, and compared to no-join, it worsens the performance for the 
ML task, because the few identified results make the joined table sparse and cause overfitting. 
Despite more records marked as match by fuzzy-join and edit-join, many of them are false 
positives. \pexeso achieves the highest F1 score with +0.019 performance gain over the runner-up. 

\mysubsection{Amazon toy product classification} 
\revise{The dataset contains 10,000 rows of toy products from Amazon.com~\cite{URL:amazon}.
The task is to predict the category from 39 classes (hobbies, office, arts, etc.) of each toy. 
We use ``product\_name'' as query column. Table~\ref{tab:effect:amazon} reports the average micro-F1 
score and the data lake record marked as match. Similar results are witnessed as we have seen in company 
classification. \pexeso perform the best, reporting +0.017 performance gain over the runner-up.}

\mysubsection{Video game sales regression} 
\revise{The dataset contains 11,493 rows of video games with attributes and sales 
information~\cite{URL:game}. The task is to predict the global sales. 
We use ``Name'' (game's name) as query column. Table~\ref{tab:effect:game} reports the average MSE 
and the data lake record marked as match. Compared to no-join, all the other methods improve the 
performance. \pexeso reports the lowest MSE and reduces it by 10\% from the runner-up.}

The above tasks show that joining tables with open data enhances the accuracy and our semantic-aware 
solution yields more gains. Note we do not join with open data blindly. It is also important to 
perform feature selection over the joined results. 

\begin{table}[t]
\centering
\caption{Parameter tuning in \pexeso.}
\resizebox{\linewidth}{!}{
\begin{tabular}{|c|c|ccg|ccg|} 
\toprule
& & \multicolumn{3}{c|}{OPEN Time (s)} & \multicolumn{3}{c|}{SWDC Time (s)} \\
$|P|$ & $m$  & index & block & block + verify & index & block & block + verify \\
\midrule
1 & 2 & 456.2 & 1.12 & 123.5 & 301.6 & 0.12 & 16.4\\
1 & 4 & 464.1 & 1.25 & 142.5 & 302.9 & 0.10 & 15.2\\  
1 & 6 & 466.9 & 1.73 & 179.2 & 315.2 & 0.19 & 15.2\\ 
1 & 8 & 458.2 & 2.12 & 196.9 & 301.9 & 0.25 & 16.0\\ 
\midrule
3 & 2 & 477.9 & 1.17 & 145.9 & 412.3 & 0.17 & 12.9\\ 
\textbf{3} & \textbf{4} & 482.6 & 1.30 & 166.1 & 421.0 & 0.15 & \textbf{12.8}\\  
3 & 6 & 481.7 & 1.25 & 89.7  & 451.9 & 0.20 & 16.3\\ 
3 & 8 & 489.7 & 1.26 & 127.6 & 507.1 & 0.22 & 21.1\\
\midrule
5 & 2 & 483.7 & 1.09 & 78.7  & 448.5 & 0.15 & 12.7\\ 
5 & 4 & 478.8 & 1.23 & 58.0  & 468.3 & 0.17 & 14.2\\  
\textbf{5} & \textbf{6} & 527.9 & 1.25 & \textbf{41.8} & 520.8 & 0.19 & 18.4\\ 
5 & 8 & 537.5 & 1.08 & 68.6 & 595.5 & 0.23 & 23.2\\ 
\midrule
7 & 2 & 579.9 & 1.18 & 95.4 & 518.0 & 0.11 & 14.8\\ 
7 & 4 & 602.6 & 1.16 & 81.0 & 568.3 & 0.13 & 15.7\\  
7 & 6 & 647.2 & 1.05 & 54.0 & 619.3 & 0.15 & 17.9\\ 
7 & 8 & 765.7 & 1.56 & 62.4 & 695.6 & 0.24 & 20.0\\ 
\midrule
9 & 2 & 788.5 & 1.13 & 74.3 & 571.0 & 0.13 & 18.0\\ 
9 & 4 & 863.0 & 1.16 & 69.8 & 610.7 & 0.11 & 18.0\\  
9 & 6 & 899.8 & 1.09 & 67.9 & 690.6 & 0.23 & 21.3\\ 
9 & 8 & 865.3 & 1.17 & 85.4 & 758.4 & 0.27 & 22.3\\ 
\bottomrule
\end{tabular}
}
\label{tab:exp:tune}
\end{table}

\newcommand{\gray}{\rowcolor[gray]{.90}}
\begin{table*}[t]
\centering
\caption{Efficiency evaluation (OPEN and SWDC are in-memory; LWDC is out-of-core; program is terminated if processing time exceeded 2 hours).}
\resizebox{\textwidth}{!}{  
\begin{tabular}{|c|c|cccg|cccg|cccg|} 
\toprule
& & \multicolumn{4}{c}{{OPEN Search Time (s)}} & \multicolumn{4}{c}{SWDC Search Time (s)} & \multicolumn{4}{c|}{LWDC Search Time (s)}\\
$T$ & $\tau$ & \ctree & \ept & \hg & \pexeso & \ctree & \ept & \hg & \pexeso & \ctreefor & \eptfor & \hgfor & \pexfor   \\

\midrule
20\% & 2\% & 678 & 710 & 75.4 & 32.5  & 678 & 691 & 130 & 9.8  & $>$ 7200 & $>$ 7200 & 3567 & 456 \\
20\% & 4\% & 656 & 794 & 88.6 & 35.9  & 778 & 739 & 131 & 10.2 & $>$ 7200 & $>$ 7200 & 4156 & 468 \\  
20\% & 6\% & 706 & 888 & 157  & 33.7 & 599 & 683 & 134 & 10.2 & $>$ 7200 & $>$ 7200 & 4678 & 475 \\ 
20\% & 8\% & 795 & 973 & 244  & 47.5 & 567 & 696 & 133 & 10.6 & $>$ 7200 & $>$ 7200 & 4532 & 474 \\ 
\midrule
40\% & 2\% & 811 & 711 & 66.7 & 33.0  & 766 & 642 & 136 & 13.6 & $>$ 7200 & $>$ 7200 & 5678 & 514 \\
40\% & 4\% & 897 & 793 & 99.5 & 44.1  & 787 & 655 & 140 & 13.6 & $>$ 7200 & $>$ 7200 & 5895 & 556 \\  
40\% & 6\% & 899 & 884 & 165  & 42.4 & 767 & 678 & 134 & 11.6 & $>$ 7200 & $>$ 7200 & 6892 & 578 \\ 
40\% & 8\% & 905 & 967 & 277  & 54.0 & 789 & 672 & 143 & 12.0 & $>$ 7200 & $>$ 7200 & 6245 & 602 \\ 
\midrule
60\% & 2\% & 867 & 704 & 74.8 & 42.2  & 677 & 577 & 137 & 12.8 & $>$ 7200 & $>$ 7200 & 5786 & 598 \\
60\% & 4\% & 913 & 796 & 106 & 52.6  & 767 & 768 & 156 & 12.5 & $>$ 7200 & $>$ 7200 & 5409 & 601 \\  
60\% & 6\% & 922 & 884 & 177  & 51.8 & 745 & 726 & 157 & 12.8 & $>$ 7200 & $>$ 7200 & 6789 & 603 \\ 
60\% & 8\% & 932 & 957 & 279  & 52.1 & 766 & 715 & 150 & 13.0 & $>$ 7200 & $>$ 7200 & $>$ 7200   & 623 \\ 
\midrule
80\% & 2\% & 910 & 712 & 81.3 & 51.5  & 776 & 809 & 138 & 13.2 & $>$ 7200 & $>$ 7200 & 6157 & 635 \\
80\% & 4\% & 898 & 780 & 108 & 53.4  & 813 & 823 & 134 & 13.4 & $>$ 7200 & $>$ 7200 & 6245 & 622 \\  
80\% & 6\% & 903 & 907 & 199  & 59.1  & 823 & 817 & 152 & 13.4 & $>$ 7200 & $>$ 7200 & $>$ 7200   & 627 \\ 
80\% & 8\% & 934 & 913 & 266  & 68.1 & 831 & 829 & 157 & 13.6 & $>$ 7200 & $>$ 7200 & $>$ 7200   & 628 \\ 
\bottomrule
\end{tabular}
}
\label{tab:exp:search}
\end{table*}

\subsection{Parameter Tuning for Efficiency}
\label{sec:para}
There are two parameters in \pexeso: $\size{P}$, the 
number of pivots, and $m$, the number of levels in the hierarchical grids. 
Table~\ref{tab:exp:tune} shows the index construction time, the blocking time, and the total 
search time (i.e., blocking and verification) with varying $\size{P}$ and $m$ on OPEN and 
SWDC. \revise{The latter two are averaged over 1,000 queries.}
The optimal parameters are $|P| = 5$ and $m = 6$ for OPEN and $|P| = 3$ and $m = 4$ for 
SWDC. \revise{We choose these parameters as the default setting.} Next we discuss the two 
parameters respectively. 

\mysubsection{Varying $\size{P}$}
When we increase the pivot size, the index construction spends more time. 
The search time first drops and then rebounds. This is because a larger pivot set filters 
more vectors but increases the number of cells in the hierarchical grids and cause more 
candidate pairs in the form of (vector, cell). 

\mysubsection{Varying $m$}
The effect of $m$ is similar to that of $\size{P}$. This is because a larger $m$ yields 
finer granularity of the hierarchical grids and improves the filtering power, 
while it results in more overhead for inverted index access. 

\mysubsection{Justification of cost analysis} 
We also evaluate the optimal $m$ obtained by our cost analysis (Section~\ref{sec:analysis}). 
The optimal $m$ obtained by analysis is 5 (4.4 before ceiling) on OPEN and 4 (3.7 before ceiling) 
on SWDC, while the empirically optimal values are 6 and 4 on the two datasets, respectively. 
This suggests that our analysis is effective in \pexeso's index construction. In addition, 
the result in Table~\ref{tab:exp:tune} shows that the blocking time is negligible in the 
overall search time, which justifies our assumption for the cost analysis. 


\subsection{Efficiency Evaluation}
\label{sec:efficiency}

\mysubsection{Performance on in-memory search}
Table~\ref{tab:exp:search} (left 2/3 part) summarizes the search time for the 
in-memory case on OPEN and SWDC, averaged over 1,000 queries. 
\pexeso performs the best in all the cases. It is 14 to 76 times faster than the 
non-blocking methods and 1.6 to 13 times faster than \hg. 

\mysubsection{Varying $\tau$}
From Table~\ref{tab:exp:search}, we also observe
the search time trends with 
varying distance threshold $\tau$. In general, the search time increases with 
$\tau$. This is because the range query condition becomes looser. 
For example, for \ctree, a larger $\tau$ causes more overlapping 
tree nodes; for \pexeso, more candidates survive the filtering of hierarchical grids. 

\mysubsection{Varying $T$}
We observe the search time generally increases with $T$ from Table~\ref{tab:exp:search}. 
The reason is that the methods are equipped with the 
early termination technique such that whenever the joinability counter reaches $T$, the 
column is immediately confirmed as joinable. 
When $T$ increases, this early termination becomes less effective and results in more 
search time. Nonetheless, \pexeso is less vulnerable to this effect due to its inverted 
index-based verification. 

\begin{figure}[t]
  \begin{subfigure}[b]{0.235\textwidth}
    \includegraphics[width=\linewidth]{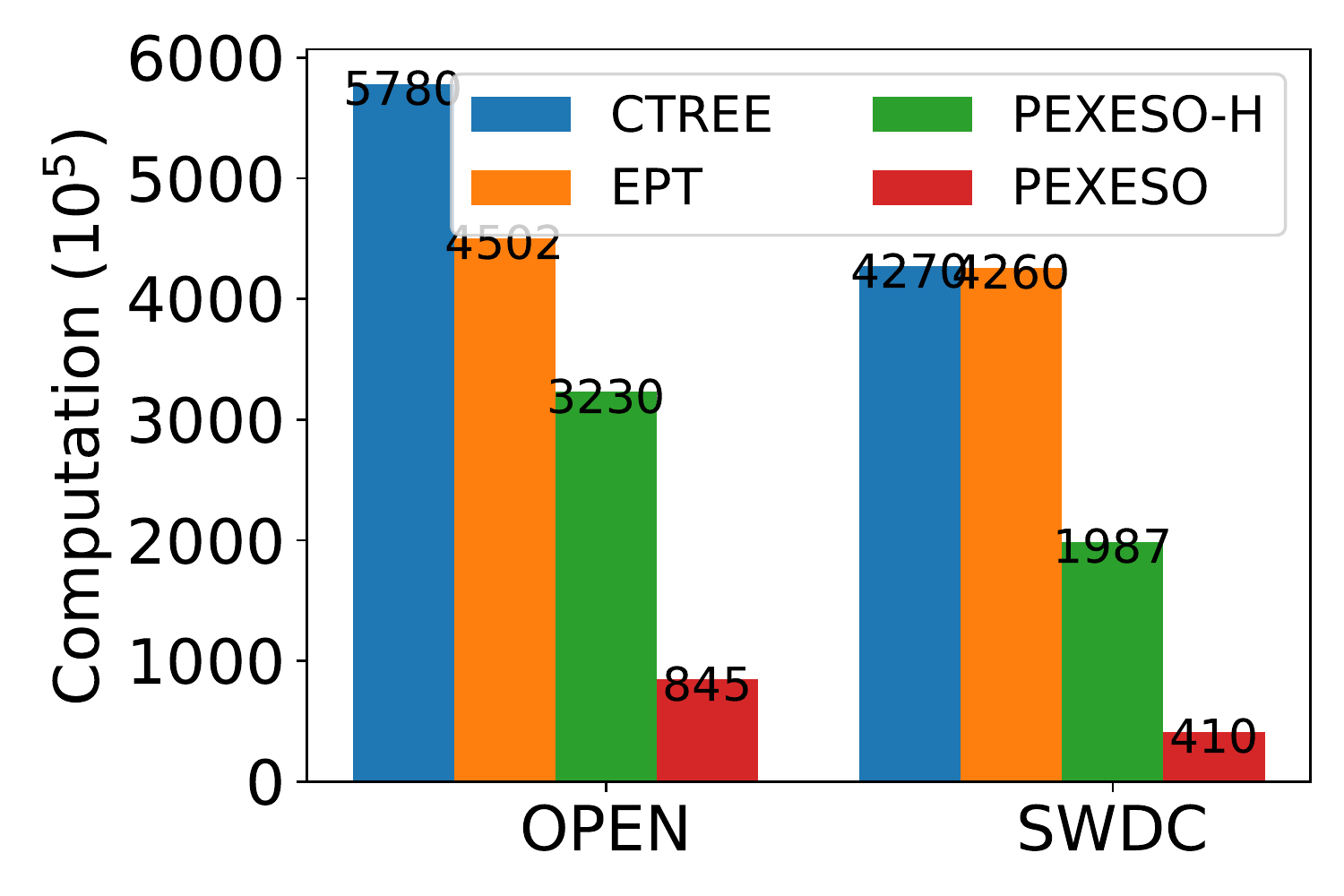}
    \caption{Distance computation times.}
    \label{exp:compute}
  \end{subfigure}
  \begin{subfigure}[b]{0.235\textwidth}
    \includegraphics[width=\linewidth]{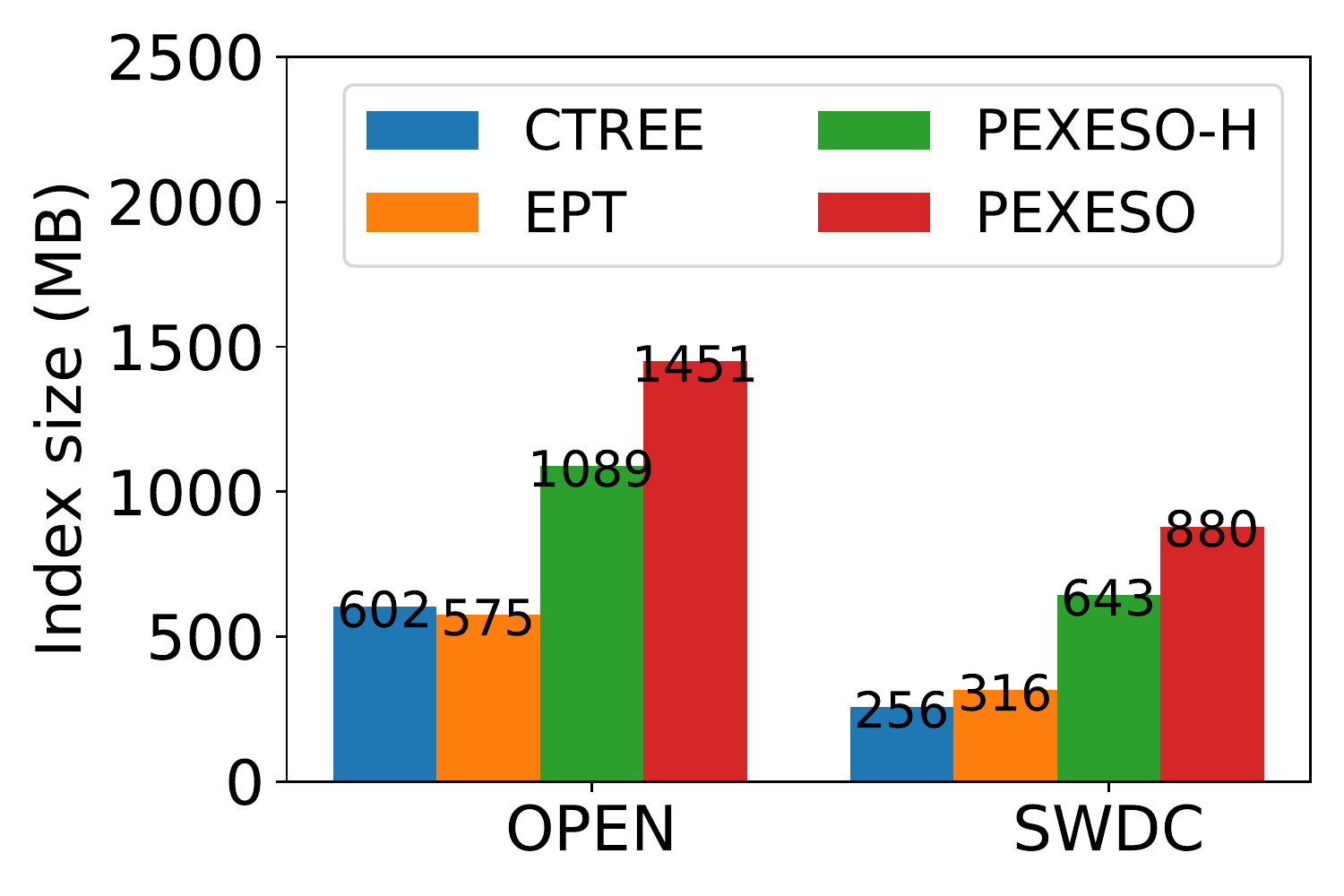}
    \caption{\revise{Index size}.}
    \label{exp:mem}
  \end{subfigure}
  \caption{Distance computation and index size (OPEN and SWDC).}
  \label{exp:comp:mem}
\end{figure}

\mysubsection{Distance computation}
To better understand why \pexeso is faster, we plot the number of distance computations 
in Fig.~\ref{exp:compute}. \pexeso reports far less times of distance computation than 
the other options. The result also shows that our blocking is useful in reducing distance 
computation, as \hg also reports less distance computation times than the other baselines.

\mysubsection{Index size}
\revise{
Fig.~\ref{exp:mem} shows the index size comparison. 
Albeit highest, the index size of \pexeso is only 2 times \ctree or \ept. Considering the 
significant speedup we have witnessed, it is worth spending moderately more space. Moreover, 
most memory consumption is the table repository storage.}

\mysubsection{Performance on out-of-core search}
We partition the LWDC dataset into 10 parts with the JSD clustering 
(Section \ref{sec:boost}). 
\revise{The index is in-memory and the dataset is disk-resident.} 
Table~\ref{tab:exp:search} (right 1/3 part) reports the search time, which includes the overhead 
of loading the data from disks. Note that we report the time only if it is within 2 hours. 
\pexeso is still the fastest, and it is 8 to 11 times faster than \hg. 

\begin{figure}[t]
  \begin{subfigure}[b]{0.235\textwidth}
    \includegraphics[width=\linewidth]{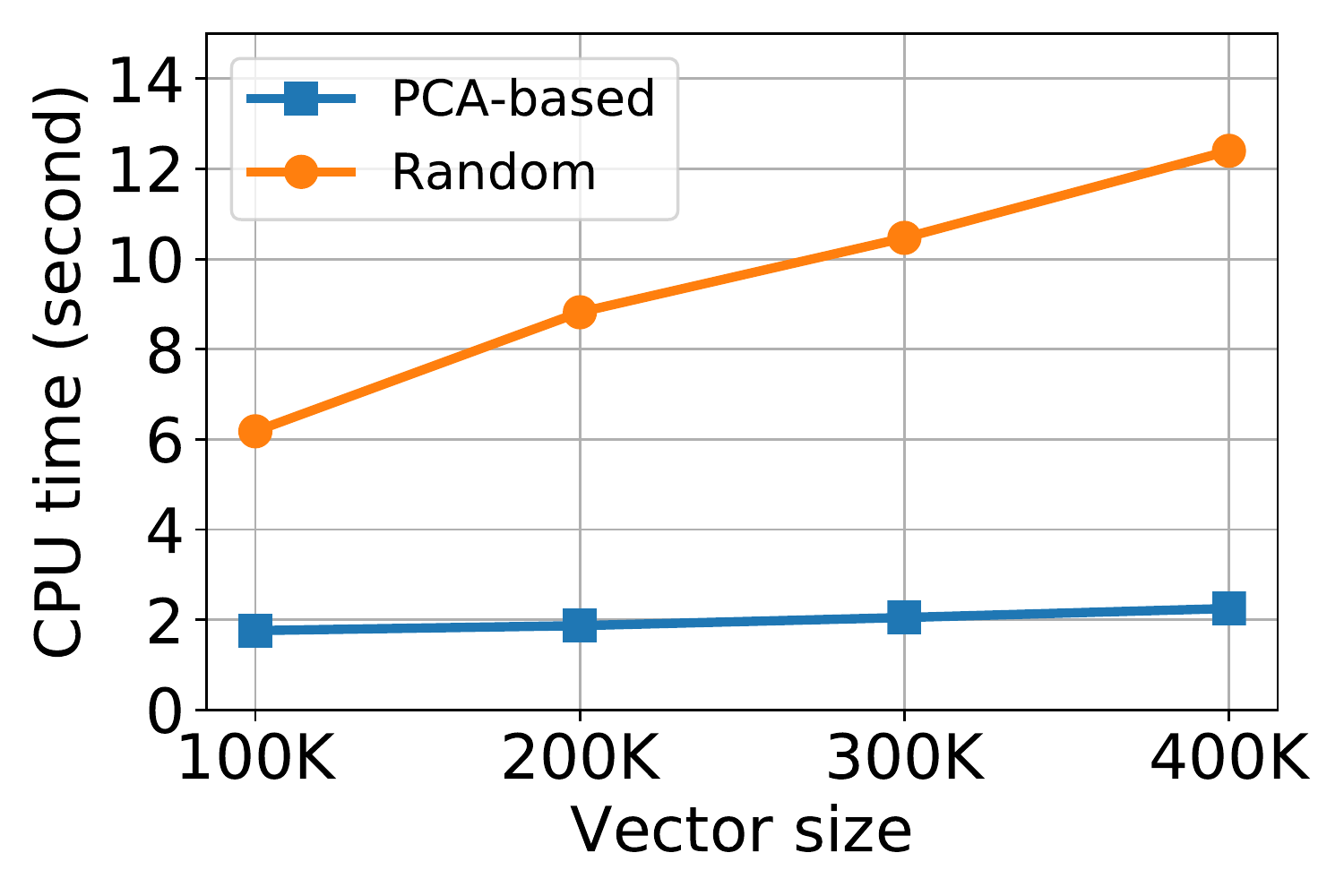}
    \caption{Pivot selection algorithms.}
    \label{exp:pivot-selection}
  \end{subfigure}  
  \begin{subfigure}[b]{0.235\textwidth}
    \includegraphics[width=\linewidth]{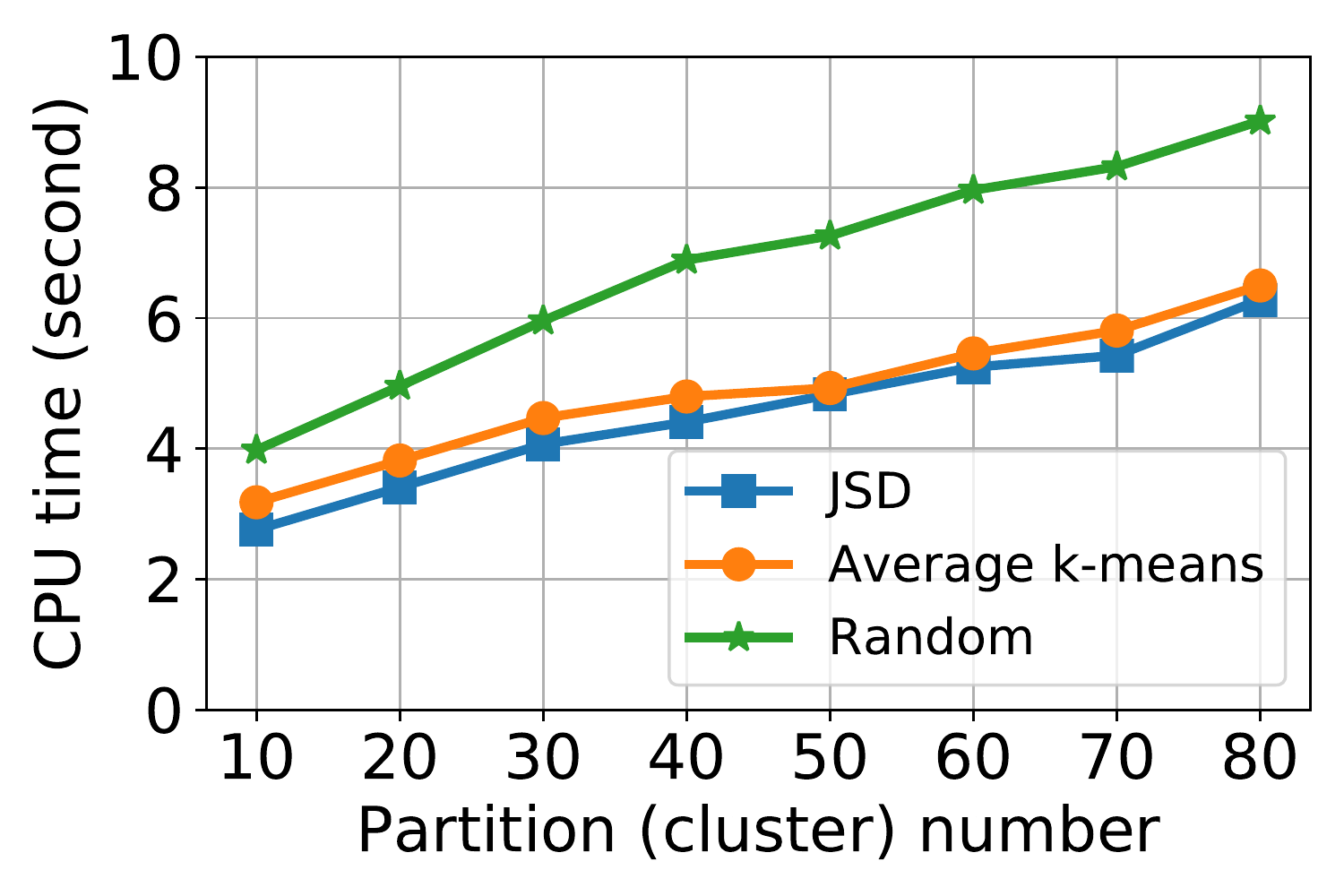}
    \caption{Data partitioning algorithms.}
    \label{exp:cluster}
  \end{subfigure}
  \caption{Pivot selection and data partitioning (LWDC).}
  \label{exp:cluster:pivot}
\end{figure}

\mysubsection{Pivot selection and data partitioning}
\revise{To show the PCA-based pivot selection is a good choice, we compare with a baseline that 
randomly selects data points from the dataset. Fig.~\ref{exp:pivot-selection} reports the 
difference. The PCA-based method is overwhelmingly better, especially when there are more vectors 
in the dataset.} 
To evaluate the proposed partitioning algorithm, we sample from LWDC 10,000 tables with 32,549 
columns. 
We partition the set of columns with the proposed JSD clustering, random partitioning, and  
average $k$-means clustering (i.e., regarding each column as the average of its vectors and 
running a $k$-means clustering). Fig.~\ref{exp:cluster} shows the search time with varying 
number of clusters. The proposed method is consistently better: it is 1.4 to 1.6 times faster 
than random partitioning and 1.1 to 1.2 times faster than average $k$-means clustering.

\begin{figure}[t]
  \begin{subfigure}[b]{0.235\textwidth}
    \includegraphics[width=\linewidth]{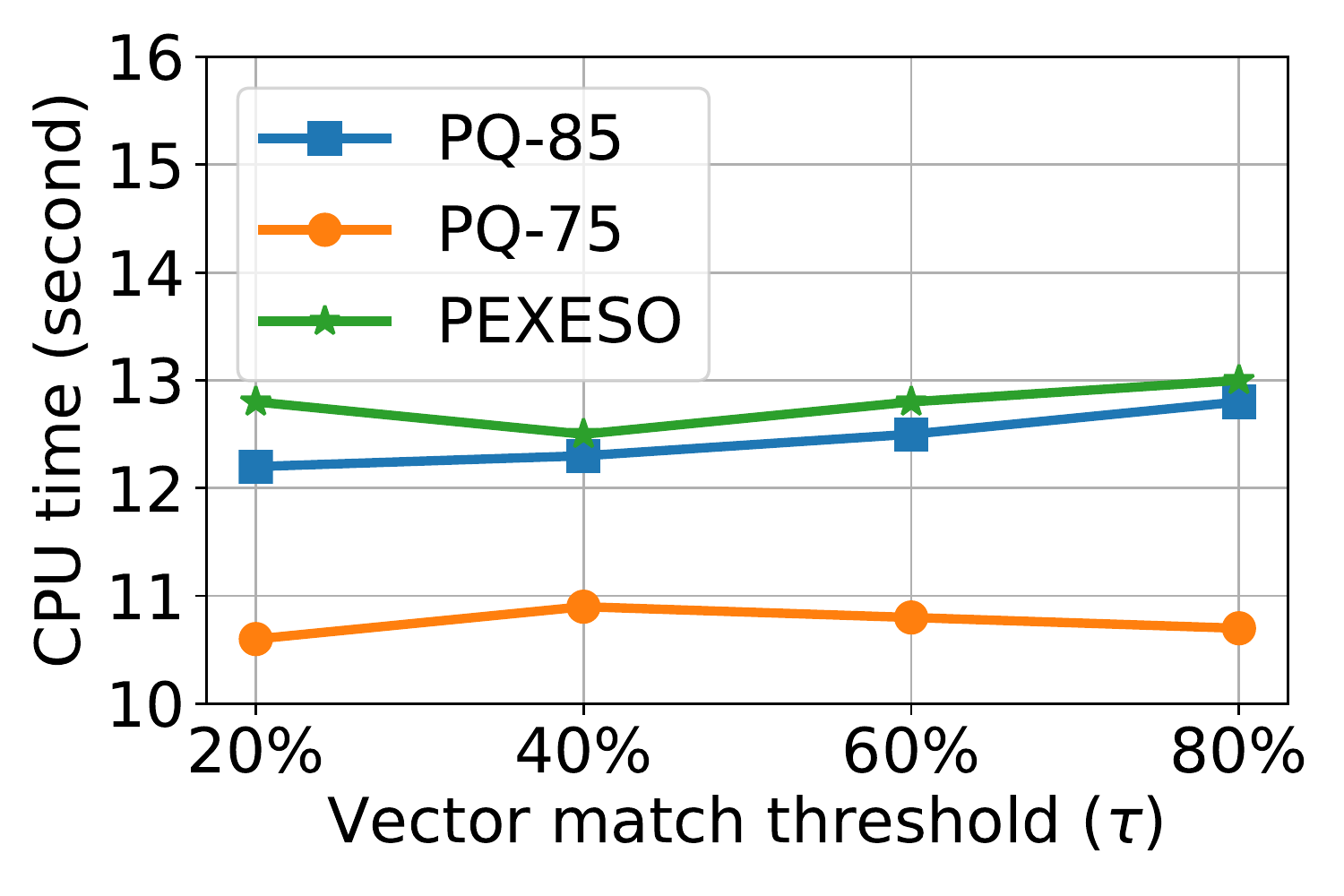}
    \caption{Varying $\tau$.}
    \label{exp:tau:pqswdc}
  \end{subfigure}
  \begin{subfigure}[b]{0.235\textwidth}
    \includegraphics[width=\linewidth]{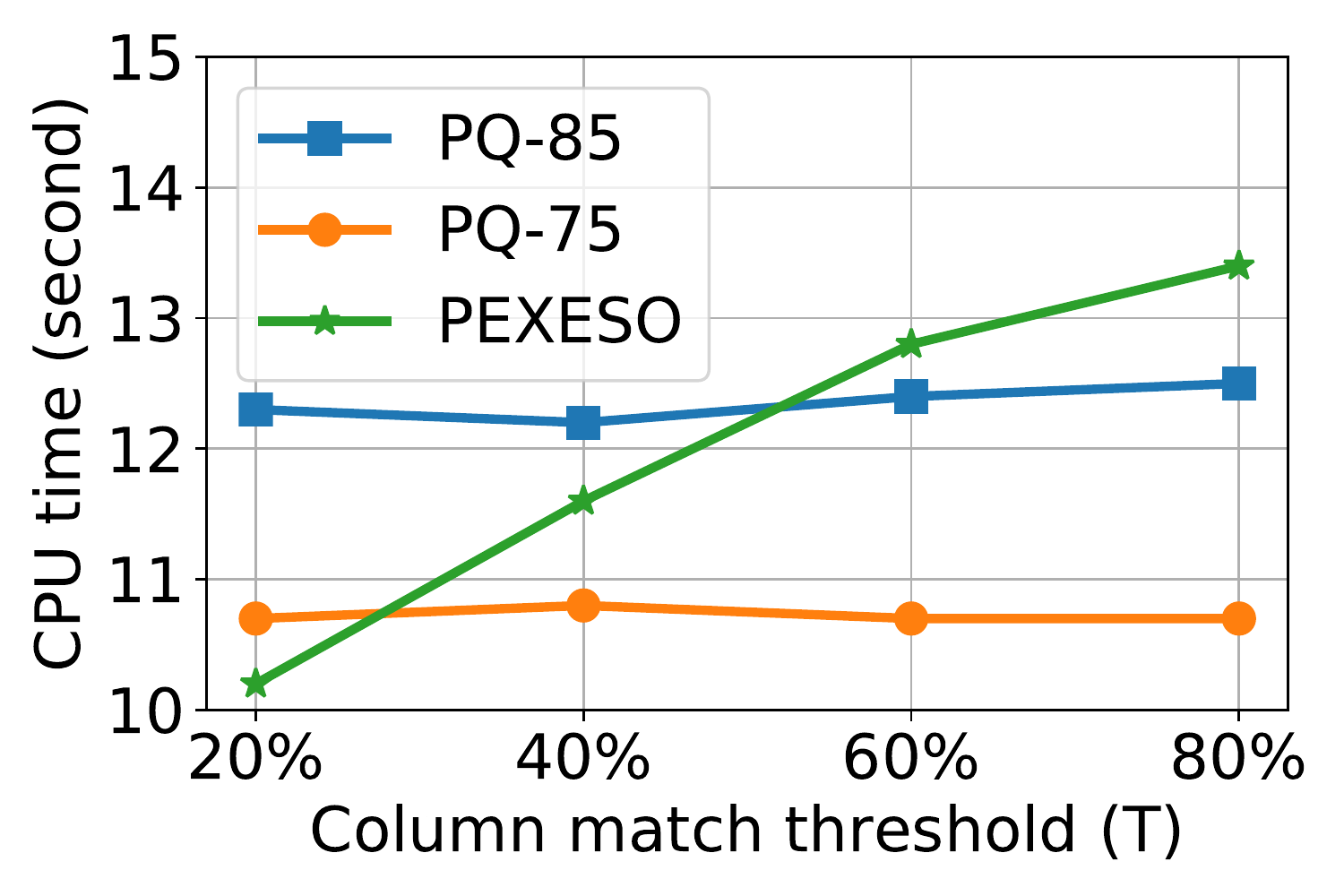}
    \caption{Varying $T$.}
    \label{exp:t:pqswdc}
  \end{subfigure}  
  \caption{Comparison to approximate method \pq (SWDC).}
  \label{exp:pq:swdc}
\end{figure}

\mysubsection{Comparison with approximate method}
We compare \pexeso with an approximate method of product quantization (\pq). 
We adjust \pq to make the recall of range query at least 75\% and 85\% and denote the 
resultant method \pq-75 and \pq-85, respectively. 
Fig.~\ref{exp:pq:swdc} plots the search time on SWDC. \pexeso is competitive with 
\pq-85, and it is even faster than \pq-75 and \pq-85 when $T$ is 20\% query column size. 

\begin{figure}[t]
  \begin{subfigure}[b]{0.235\textwidth}
    \includegraphics[width=\linewidth]{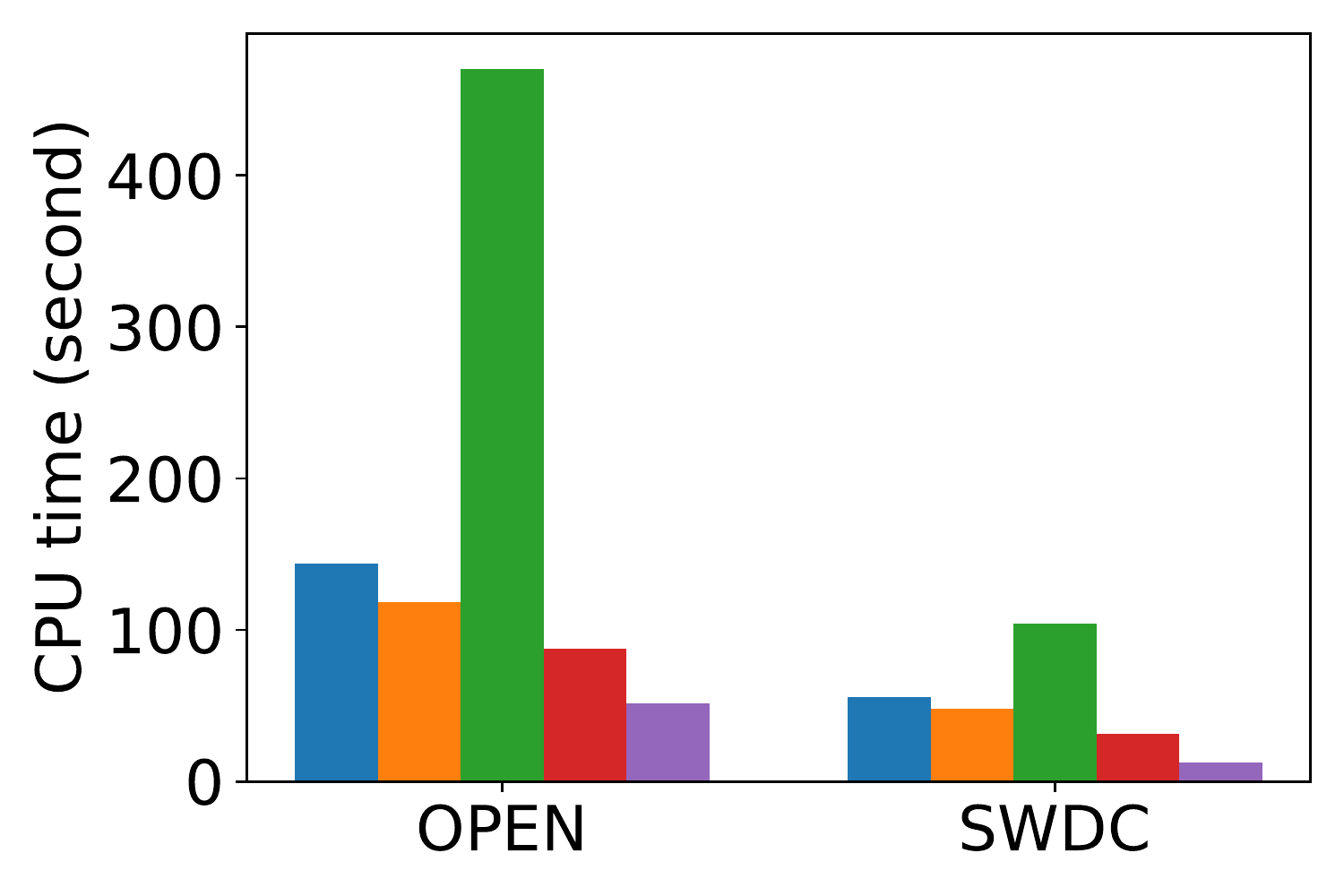}
    \label{exp:ablation-open-swdc}
  \end{subfigure}
  \begin{subfigure}[b]{0.235\textwidth}
    \includegraphics[width=\linewidth]{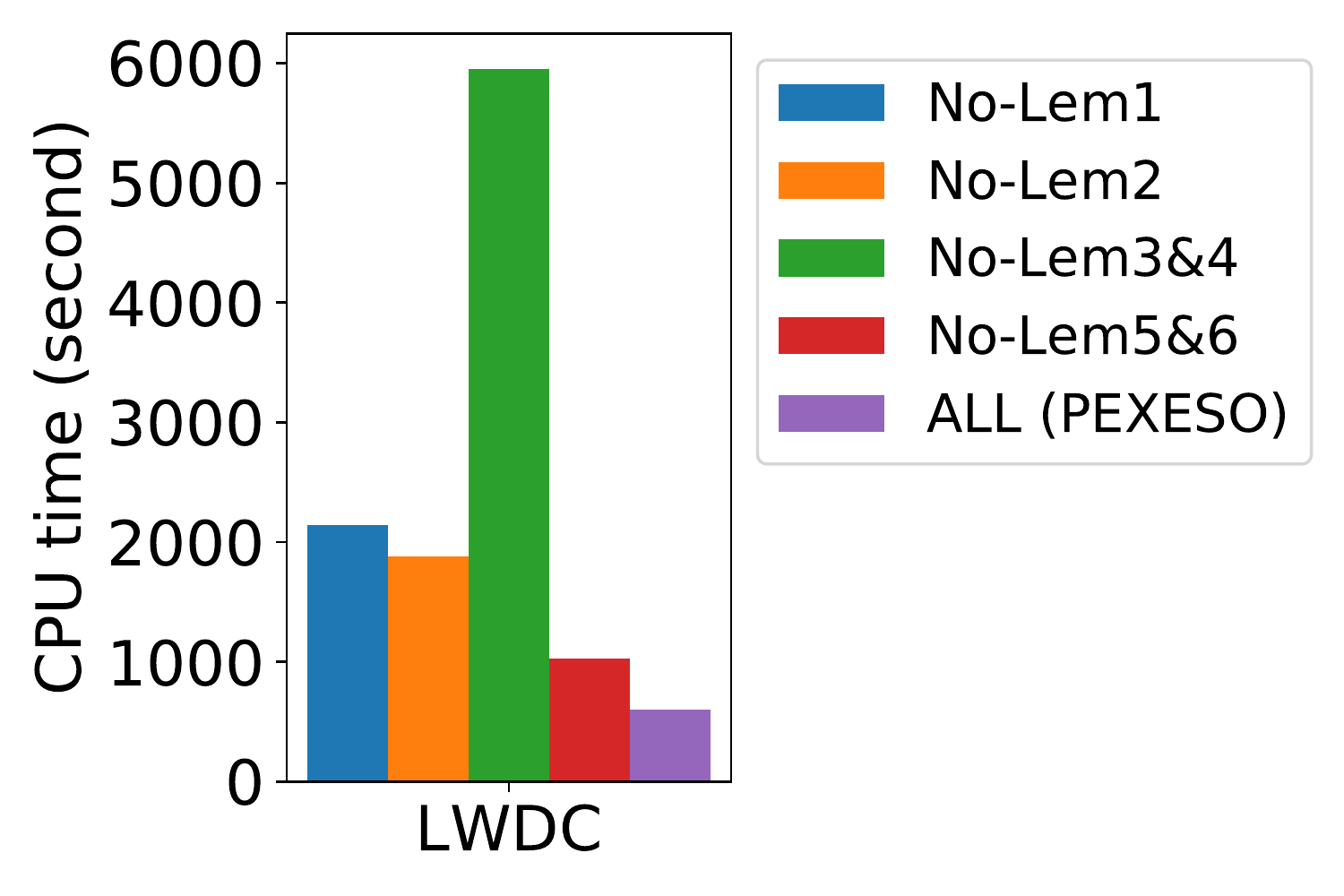}
    \label{exp:ablation-lwdc}
  \end{subfigure}  
  \caption{\revise{Ablation study (OPEN, SWDC, and LWDC)}.}
  \label{exp:ablation}
\end{figure}

\mysubsection{Ablation study}
\revise{We divide the lemmata into four groups and remove each group at a time. 
Fig.~\ref{exp:ablation} shows how they affect search time. The filtering ones 
(Lemmata~\ref{lem:pivot}, \ref{lem:vcfilter}, and \ref{lem:ccfilter}) 
are more effective than their matching counterparts (Lemmata~\ref{lem:match}, 
\ref{lem:vcmatch}, and \ref{lem:ccmatch}). The group of vector-cell 
(Lemma~\ref{lem:vcfilter}) and cell-cell (Lemma~\ref{lem:ccfilter}) filtering is 
by far the most effective. This suggests that our filtering principles developed 
upon hierarchical grids are more effective than the point-wise ones used in existing 
work~\cite{pviotvldb17}.}

\begin{figure*}[t]
  \centering
  \begin{subfigure}[b]{0.18\textwidth}
    \includegraphics[width=\linewidth]{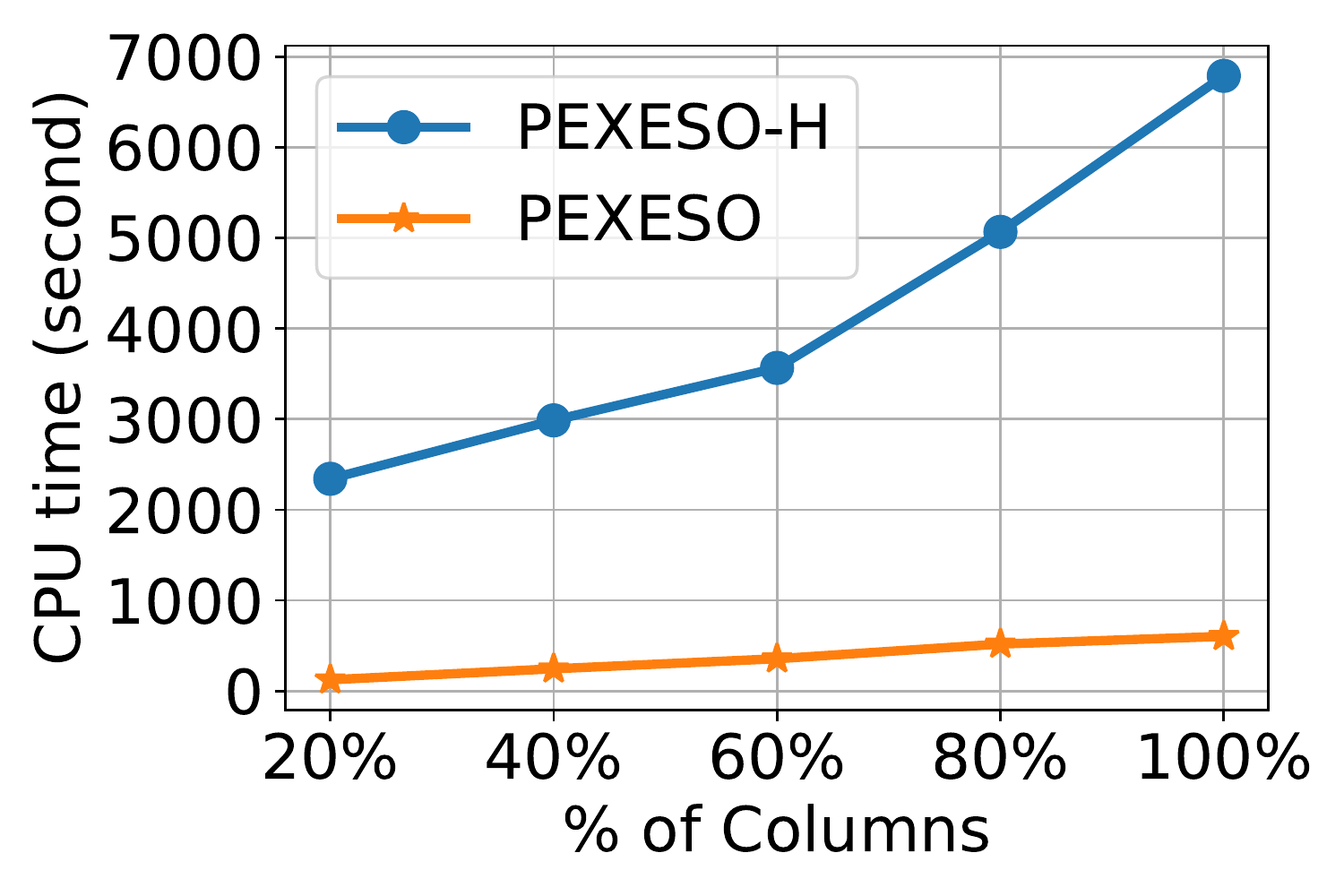}
    \caption{Varying \% of columns.}
    \label{exp:col:lwdc}
  \end{subfigure}
  \begin{subfigure}[b]{0.18\textwidth}
    \includegraphics[width=\linewidth]{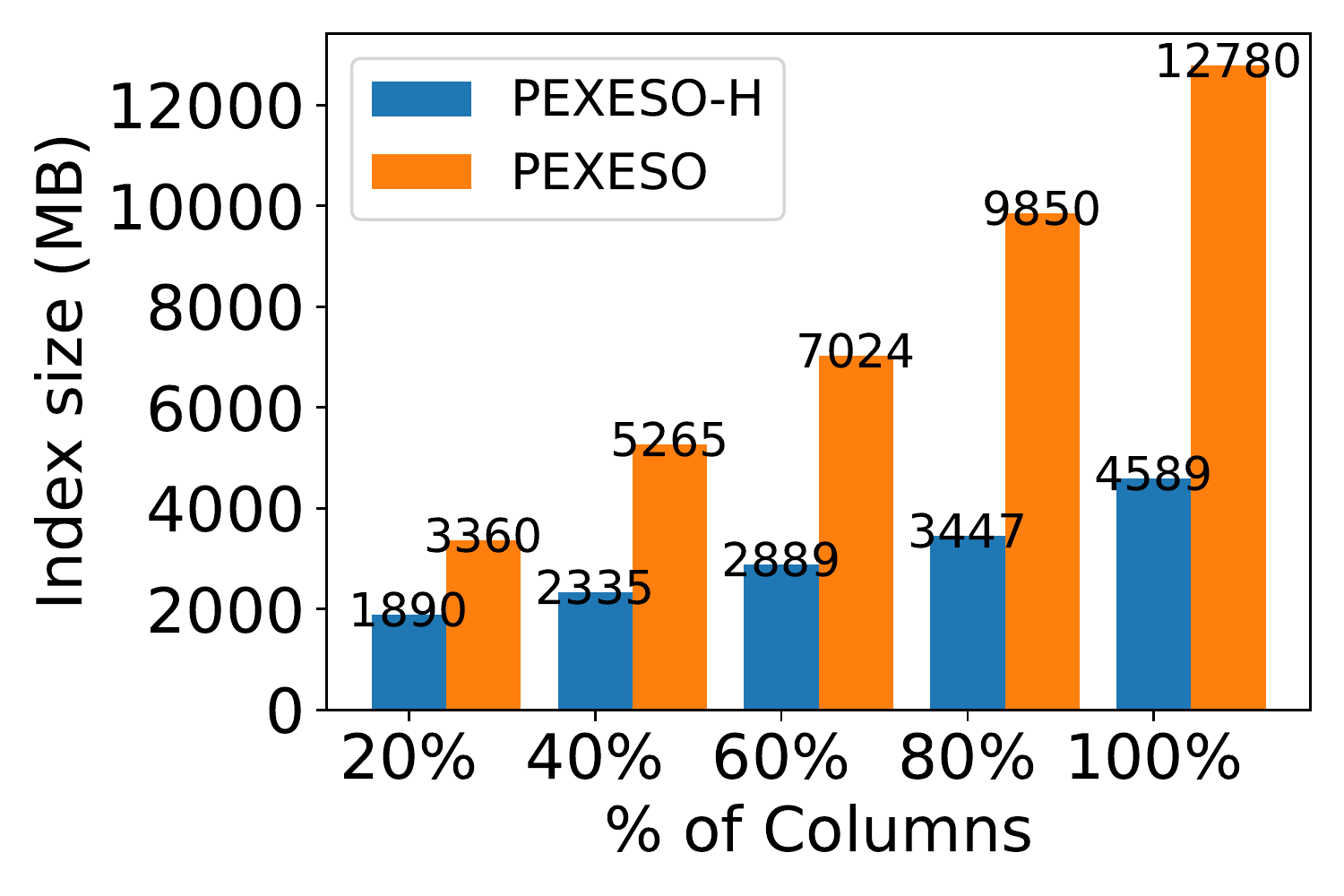}
    \caption{Varying  \% of columns.}
    \label{exp:col:index-size-col:lwdc}
  \end{subfigure}
  \begin{subfigure}[b]{0.18\textwidth}
    \includegraphics[width=\linewidth]{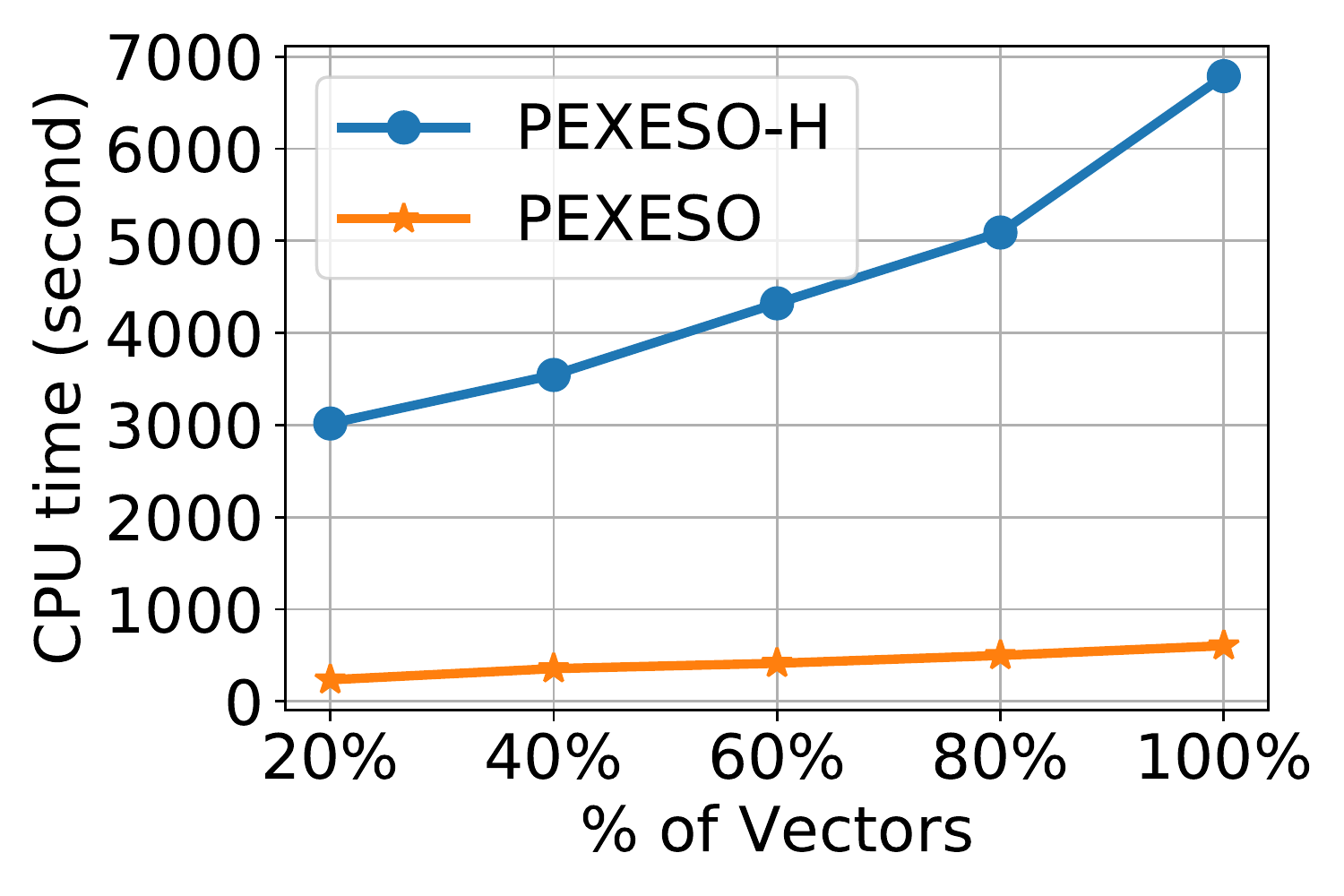}
    \caption{Varying \% of vectors.}
    \label{exp:vec:lwdc}
  \end{subfigure}
  \begin{subfigure}[b]{0.18\textwidth}
    \includegraphics[width=\linewidth]{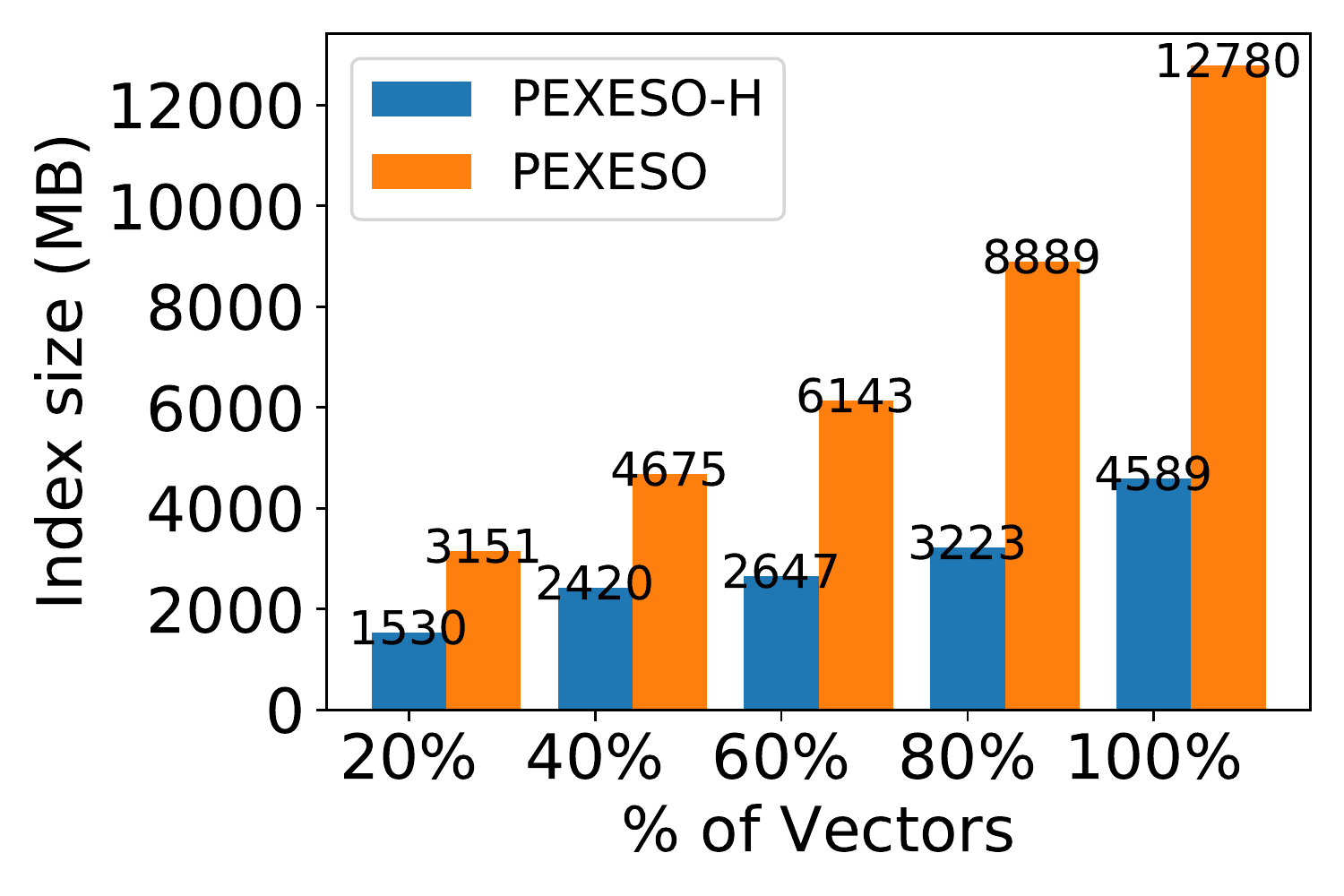}
    \caption{Varying  \% of vectors.}
    \label{exp:col:index-size-vec:lwdc}
  \end{subfigure}
  \begin{subfigure}[b]{0.18\textwidth}
    \includegraphics[width=\linewidth]{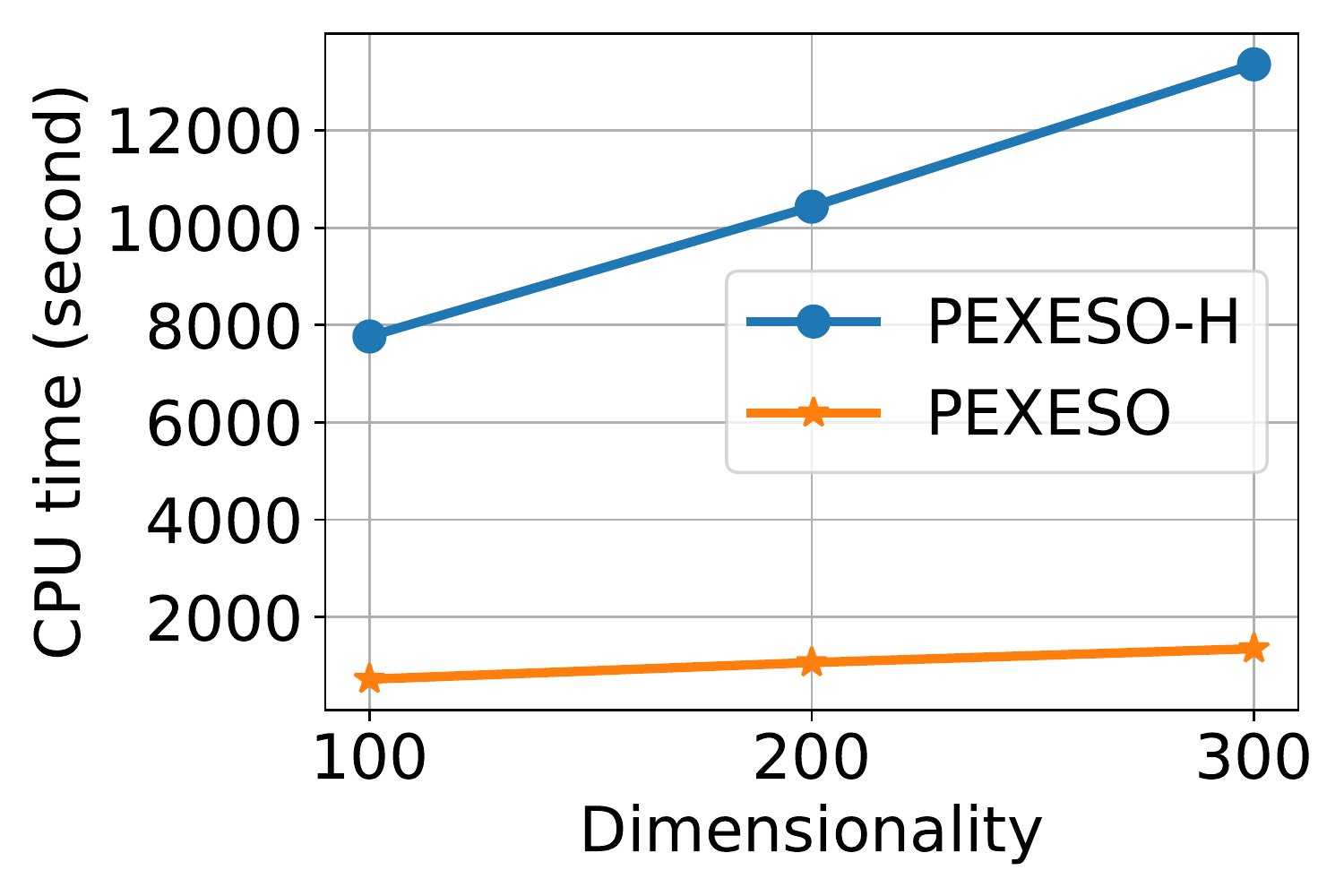}
    \caption{Varying dimensinality.}
    \label{exp:dim:lwdc}
  \end{subfigure}  
  \caption{Scalability evaluation (LWDC).}
  \label{exp:scale:lwdc}
\end{figure*}

\mysubsection{Scalability evaluation}
\revise{We vary the number of columns, the number of vectors, and the dimensionality of embeddings to 
evaluate the scalability of \pexeso. To vary the number of columns, we uniformly sample columns 
from the table repository. To vary the number of vectors, we do not sample from the collection 
of vectors but uniformly sample a percentage of rows from each column. Figs.~\ref{exp:col:lwdc} 
--~\ref{exp:dim:lwdc} plot the search time and index size on out-of-core LWDC, where only \pexeso 
and \hg are shown because the other methods are too slow. When varying the number of columns or 
vectors, \pexeso's search time and index size scale almost linearly while \hg reports superlinear 
growth. The reason is \pexeso utilizes an inverted index-based verification technique which 
reduces the number of vector pairs for distance computation to almost linear. The search times of 
both methods scale almost linearly with the dimensionality of embeddings. This is because the 
distance computation is linear in the dimensionality, and it dominates the overall search time. 
Their index sizes do not change with the dimensionality because they are constructed for the pivot 
space.}

\section{Related Work} \label{sec:related}

\mysubsection{Related table discovery}
Besides joinable table search~\cite{josie, lshensemble}, there are studies 
on finding related tables with criteria \revise{other than joinability}. 
Nargesian \etal \cite{lsh-table-union} developed LSH-based techniques for 
searching unionable tables, \revise{i.e., searching in a data lake for 
columns that can be vertically concatenate to a query table}. 
Zhang and Ives studied related table discovery with a composite score of 
multiple similarities~\cite{DBLP:conf/sigmod/ZhangI20}. Bogatu \etal 
\cite{datadiscovery} also proposed a scoring function involving multiple 
attributes of a table and studied on finding \topk results. 

\mysubsection{Similarity in metric space} 
There has been plenty of work in this area. We refer readers to \cite{high-dimensional-tutorial} 
for a recent survey. The most related ones to our work are: 
Yu \etal \cite{iDistance-join} applied the iDistance technique to \knn join. Fredriksson and 
Braithwaite~\cite{quicker-join} improved the quick join algorithm for similarity joins. 
Pivot-based methods were surveyed in \cite{pviotvldb17, spbtreeTKDE17}. 
\revise{These methods answer similarity queries but cannot solve our problem efficiently for the 
following reasons: 
\begin{inparaenum} [(1)]
  \item the indexing methods rebuild the index when the threshold changes and they also need 
  an index for the query column, and 
  \item the non-indexing methods deal with one-time joins, whereas joinable table 
  search may be invoked multiple times in a data lake. 
\end{inparaenum}
We utilize a hierarchical grid to partition the pivot space and develop filtering 
conditions and verification techniques specific to our problem.} 

\mysubsection{Joining related tables} 
Set and string similarities have also been used to find related tables. For query 
processing algorithms, we refer readers to \cite{set-sim-join-survey-vldb16} for 
experimental comparison and \cite{pigeonring} for recent advances. 
Wang \etal designed a fuzzy join predicate that combines token and characters and 
proposed the corresponding algorithm~\cite{fuzzytods}. Deng \etal \cite{silkmoth} 
studied the related set (table) search problem that finds sets with the maximum 
bipartite matching metrics. Wang \etal proposed MF-join \cite{mf-join-icde19} that 
performs a fuzzy match with multi-level filtering. 
The above solutions were not designed for data lakes (see \cite{josie}) 
and only deal with raw textual data \revise{rather than high-dimensional vectors}. 
\revise{Zhu \etal proposed auto-join~\cite{auto-join}, which joins two tables with 
string transformations on columns. He \etal proposed SEMA-join~\cite{sema-join}, 
which finds related pairs between two tables with statistical correlation. Although
optimizations for joining two given tables were introduced in the two studies, when 
applying to our problem, it is prohibitive to try joining every table in the data 
lake with the query table.}

\section{Conclusion} \label{sec:concl}
We studied the problem of joinable table discovery in data lakes. We 
proposed the \pexeso framework which utilizes pre-trained models to 
transform textual attributes to high-dimensional vectors so that 
records can be semantically joined via similarity predicates and more 
meaningful results can be identified. To speed up the search process, 
we designed an indexing method along with a block-and-verify algorithm 
based on pivot-based filtering. We proposed a partitioning method to 
handle the out-of-core case for very large data lakes. The experiments 
showed that \pexeso outperforms alternative solutions in finding 
joinable tables, and the identified tables improve the performance of 
building ML models. The experiments also demonstrated the superiority 
of \pexeso in efficiency. 


\section*{Acknowledgement}
We thank Dr. Genki Kusano and Dr. Takuma Nozawa (NEC Corporation) for discussions.


\bibliographystyle{abbrv}
\bibliography{main}  





\end{document}